\documentclass[CRGEOS,Unicode,manuscript]{cedram}

\usepackage{xspace}
\usepackage{subcaption}
\usepackage{comment}
\usepackage{bm}
\renewcommand{\thetable}{\Roman{table}}%

\renewcommand{\Pr}{\mathrm{Pr}}
\renewcommand{\Re}{\mathrm{Re}}
\newcommand{\tauell}{$\tau$-$\ell$\xspace}
\newcommand{\Ek}{\mathrm{Ek}}
\newcommand{\Ro}{\mathrm{Ro}}
\newcommand{\Roff}{\mathrm{Ro}_{\mspace{-1mu}f\mspace{-4mu}f}}
\newcommand{\Rm}{\mathrm{Rm}}
\newcommand{\Pm}{\mathrm{Pm}}
\newcommand{\Ra}{\mathrm{Ra}}
\newcommand{\Lu}{\mathrm{Lu}}

\newcommand{\sslash}{\mathbin{/\mkern-4mu/}}

\title[\tauell regime diagrams]{Dynamic regimes in planetary cores: \tauell diagrams}

\author{\firstname{Henri-Claude} \lastname{Nataf}\CDRorcid{0000-0001-6737-1314}\IsCorresp}
\address{Univ. Grenoble Alpes, Univ. Savoie Mont Blanc, CNRS, IRD, Univ. Gustave Eiffel, ISTerre, 38000 Grenoble, France}
\email[H-C. Nataf]{Henri-Claude.Nataf@univ-grenoble-alpes.fr}
\author{\firstname{Nathana\"el} \lastname{Schaeffer}\CDRorcid{0000-0001-5206-3394}}
\addressSameAs{1}{Univ. Grenoble Alpes, Univ. Savoie Mont Blanc, CNRS, IRD, Univ. Gustave Eiffel, ISTerre, 38000 Grenoble, France}
\email[N. Schaeffer]{Nathanael.Schaeffer@univ-grenoble-alpes.fr}


\begin{abstract} 
Planetary cores are the seat of rich and complex fluid dynamics, in which the effects of rotation and magnetic field combine.
The equilibria governing the strength of the magnetic field produced by the dynamo effect, the organisation and amplitude of the flow, and those of the density field, remain debated despite remarkable progress made in their numerical simulation.
This paper describes an approach based on the explicit consideration of the variation of time scales $\tau$ with spatial scales $\ell$ for the different physical phenomena involved. 
The \tauell diagrams thus constructed constitute a very complete graphic summary of the dynamics of the object under study.
We highlight the role of the available convective power in controlling this dynamics, together with the relevant force balance, for which we derive a very telling \tauell translation.
Several scenarios are constructed and discussed for the Earth's core, shedding new light on the width of convective columns, and on the force equilibria to be considered.
A QG-MAC scenario adapted from \citet{aubert2019} gives a good account of the observations.
A diversion to Venus reveals the subtlety and relativity of the notion of 'fast rotator'.
We discuss scaling laws and their validity domain, and illustrate `path strategies'.
A complete toolbox is provided, allowing everyone to construct a \tauell diagram of a numerical simulation, a laboratory experiment, a theory, or a natural object.

Supplementary material for this article is supplied as a separate 
archive \cdrattach{Nataf\_Schaeffer\_SupMat.zip}, the related data is displayed in
document \cdrattach{Nataf\_Schaeffer\_SupMat.pdf}.
\end{abstract}

\alttitle{R\'egimes dynamiques dans les noyaux plan\'etaires : diagrammes \tauell}
\begin{altabstract}
Les noyaux plan\'etaires sont le si\`ege d'une dynamique des fluides riche et complexe o\`u se combinent les effets de la rotation et du champ magn\'etique.
Les \'equilibres gouvernant l'intensit\'e du champ magn\'etique produit par effet dynamo, l'organisation et l'amplitude de l'\'ecoulement, et celles du champ de densit\'e, demeurent d\'ebattus, malgr\'e les progr\`es remarquables de leur simulation num\'erique.
Cet article d\'etaille une approche qui repose sur la prise en compte explicite de la variation des \'echelles de temps $\tau$ avec les \'echelles spatiales $\ell$ pour les diff\'erents ph\'enom\`enes physiques impliqu\'es.
Les diagrammes \tauell ainsi construits constituent un r\'esum\'e graphique tr\`es complet de la dynamique de l'objet \'etudi\'e.
Nous soulignons le r\^ole que joue la puissance convective disponible dans l'\'etablissement de cette dynamique, ainsi que l'\'equilibre des forces pour lequel nous d\'erivons une traduction \tauell qui se  r\'ev\`ele tr\`es parlante.
Plusieurs scenarios sont ainsi construits et discut\'es pour le noyau terrestre, apportant un
nouvel \'eclairage sur la largeur des colonnes convectives, et sur les \'equilibres de force \`a consid\'erer.
Un scenario QG-MAC adapt\'e de \citet{aubert2019} rend bien compte des observations.
Un d\'etour par V\'enus r\'ev\`ele la subtilit\'e et la relativit\'e de la notion de `rotateur rapide'.
Nous discutons les lois d'\'echelle et leur domaine de validit\'e, et illustrons le concept de `path strategy`.
Une bo\^ite \`a outils compl\`ete est fournie, permettant \`a chacun de construire le diagramme \tauell d'une simulation num\'erique, d'une exp\'erience de laboratoire, d'une th\'eorie, ou d'un objet naturel.
\end{altabstract}

\begin{document}

\maketitle


\section{Introduction}
\label{sec:introduction}

Enormous progress has been achieved in the modeling and understanding of the magnetic dynamo at work in the core of the Earth and other planets since the first 3D numerical simulations of \citet{glatzmaier1995} and \citet{kageyama1995}.
It rapidly appeared that the magnetic fields produced by such numerical simulations met the main characteristics of the long-term magnetic field observed on Earth, such as its dipolarity, the presence of high-flux patches at high latitudes, and symmetry properties \citep{christensen1999,olson2006,christensen2010a}.
Magnetic intensity scaling laws for planetary and stellar dynamos were obtained by combining an analysis of the dominant terms of the governing equations with results of an extensive survey of numerical simulations \citep{christensen2006,christensen2010b}.

In the meantime, shorter timescale manifestations of the geodynamo were unveiled, such as a large-scale off-centered anticyclone \citep[\textit{e.g.},][]{pais2008}, and torsional waves (geostrophic Alfv\'en waves) with periods of a few years \citep{gillet2010}.
These new observations prompted efforts to run numerical simulations at more extreme parameter values \citep{schaeffer2017,aubert2017,aubert2023}, increasing the role of rotation by decreasing the Ekman number down to $\Ek=10^{-7}$, and increasing the convective forcing up to $\Ra/\Ra_c=6300$, where $\Ra$ is the Rayleigh number, and $\Ra_c$ its critical value.
These extreme simulations of the geodynamo successfully account for fast dynamics retrieved from observations.

In view of this remarkable progress, it might seem that most problems are solved.
In fact, hot debates are still roaming on several crucial issues.
One of them concerns the dominant length-scale of convective structures in Earth's core.
Column widths of $100$ m are suggested by \citet{yan2022} while \citet{guervilly2019} advocate $30$ km.
Extrapolating force-balances from numerical simulations and laboratory experiments to natural systems is another issue \citep{aurnou2017,schwaiger2019,teed2023}.
The relevance of scenarios with weak and strong magnetic field branches is also hotly debated \citep{dormy2016}.
One extreme viewpoint being expressed by \citet{cattaneo2022} who claim that Earth would not have been able to produce a magnetic field as strong as today without Moon's help. 

There is room for such diverging views because the distance from numerically accessible parameters to expected planetary values remains vertiginous.
Laboratory experiments somewhat enlarge the accessible range but are limited to non-dynamo regimes, making the link with numerics and observations difficult.

This is the motivation for exploring a different route: instead of extrapolating available simulations to core conditions, start from the actual expected properties of the core, and patch scenarios of turbulence that correspond to different regimes encountered at different scales.
This leads to the construction of \tauell regime diagrams of turbulence, as briefly introduced by \citet{nataf2015}.

Our experience is that this approach is an excellent intuition-booster.
It provides a simple graphical support that can greatly help deciphering and testing more mathematically-motivated approaches.
However, we observe that it has not yet received an audience, perhaps because it clearly advocates a `fuzzy physics' method, and also because it was originally published in a limited-access collection.

In this article, we present a largely renewed and extended version of the \tauell regime diagrams we originally proposed.
We detail the steps for constructing such diagrams, providing examples of application to numerical simulations and laboratory experiments.
Key properties of \tauell diagrams are highlighted and illustrated by simple examples.

The central part of the article is devoted to an application to the Earth's core.
We propose scenarios for a non-magnetic rotating convective core, and for a dynamo-generating rotating convective core.
The resulting diagrams are compared with the predictions of several scaling analyses \citep{christensen2006, christensen2010b, davidson2013, aubert2017}.

\citet{nataf2015} were building scenarios from the observed large-scale flow and magnetic field, and testing how they were compatible with the expected available power.
In this article, our \tauell diagrams are constructed to satisfy a given constraint on the convective or dissipated power, a key property of turbulent flows.
Comparison with the observed flow and magnetic field (when available) is used as a validation test.
This is a more challenging exercise, which leads us to consider the various force balances that could govern the dynamics of the object.
We derive the \tauell translation of the main relevant force balances (CIA, QG-CIA, MAC, QG-MAC, IMAC).
It turns out that these translations and their graphical representations are very telling.

This should facilitate the construction of \tauell diagrams for planets, exoplanets and stars for which no direct observation of the large-scale flow velocity and magnetic field is available.
Indeed, planets are thermal machines and their thermal evolution is probably what we can estimate best.
Liquid cores of planets cool down on geological timescales, generating convective motions.
Convective power, which can be estimated from the planet's thermal history \citep[\textit{e.g.,}][]{stevenson1983,lister2003,nimmo2015,landeau2022,driscoll2023}, sustains fluid flow and magnetic field.
Dissipation of this power by either momentum or magnetic diffusion, or both, controls the regimes of turbulence the system experiences.

We present and illustrate the construction rules and key properties of \tauell regime diagrams of turbulence in Section \ref{sec:construction}.
Section \ref{sec:phenomena} introduces the physical phenomena at work in planetary cores, and relates \tauell diagrams to classical dimensionless numbers.
Section \ref{sec:non-magnetic} presents \tauell regime diagrams for a non-magnetic rotating core.
\tauell diagrams of the present-day geodynamo are built in Section \ref{sec:Earth}.
Both sections emphasize the crucial role of the available convective power and force balances.
The discussion section \ref{sec:discussion} illustrates how \tauell diagrams bring a new light on several ongoing debates.
Limitations and perspectives are outlined in Section \ref{sec:limitations}, and we conclude in Section \ref{sec:conclusion}.
Appendix \ref{sec:spectra} provides rules for converting spectra into \tauell language.
Simple Python programs used to build \tauell diagrams are given as supplementary material, together with examples from numerical simulations and laboratory experiments.

\section{Construction rules and key properties of \tauell diagrams}
\label{sec:construction}

This section presents the rules used to construct \tauell diagrams.
Turbulent systems display a wide range of length-scales and timescales.
Timescales of physical phenomena such as diffusion or wave propagation depend upon the length-scale at which they operate.
For example, timescale $\tau_\nu$ of momentum diffusion at length-scale $\ell$ can be written as $\tau_\nu(\ell) = \ell^2/\nu$, where $\nu$ is the kinematic viscosity.
Similarly, turnover time $\tau_u$ of a vortex of radius $\ell$ is given by $\tau_u(\ell) = \ell/u(\ell)$, where $u(\ell)$ is the vortex fluid velocity.
We build \tauell regime diagrams by plotting timescales $\tau_x$ as a function of length-scale $\ell$ in a log-log plot, for all the different physical phenomena $x$ that govern the fluid flow in a given system.\\

\noindent
\fbox{%
\begin{minipage}{0.95\columnwidth}
	\textbf{Construction rules of \tauell regime diagrams}\\
    
	\tauell regime diagrams are `object-oriented'. They are built following these steps:
	
    \begin{itemize}
	\item Identify physical phenomena that play an important role in the object under study.
	\item Document relevant physical properties (viscosity, thermal diffusivity, rotation rate, etc).
	\item Build and draw lines $\tau(\ell)$ that control dissipative and wave propagation phenomena.
	\item Identify different turbulence regimes the object might experience.
	\item Construct and draw lines $\tau(\ell)$ of fields  (velocity, buoyancy, magnetic field) that describe the object's turbulent behaviour, given a dissipated power $\mathcal{P}_{diss}$.
	\item Compare predictions with observables such as large-scale flow velocity and magnetic field, when available.
	\end{itemize}
	
\end{minipage}
}


\subsection{A simple example: Kolmogorov's universal turbulence}
\label{sec:K41}
\begin{centering}
	\begin{figure}
		\includegraphics[width=7cm]{ 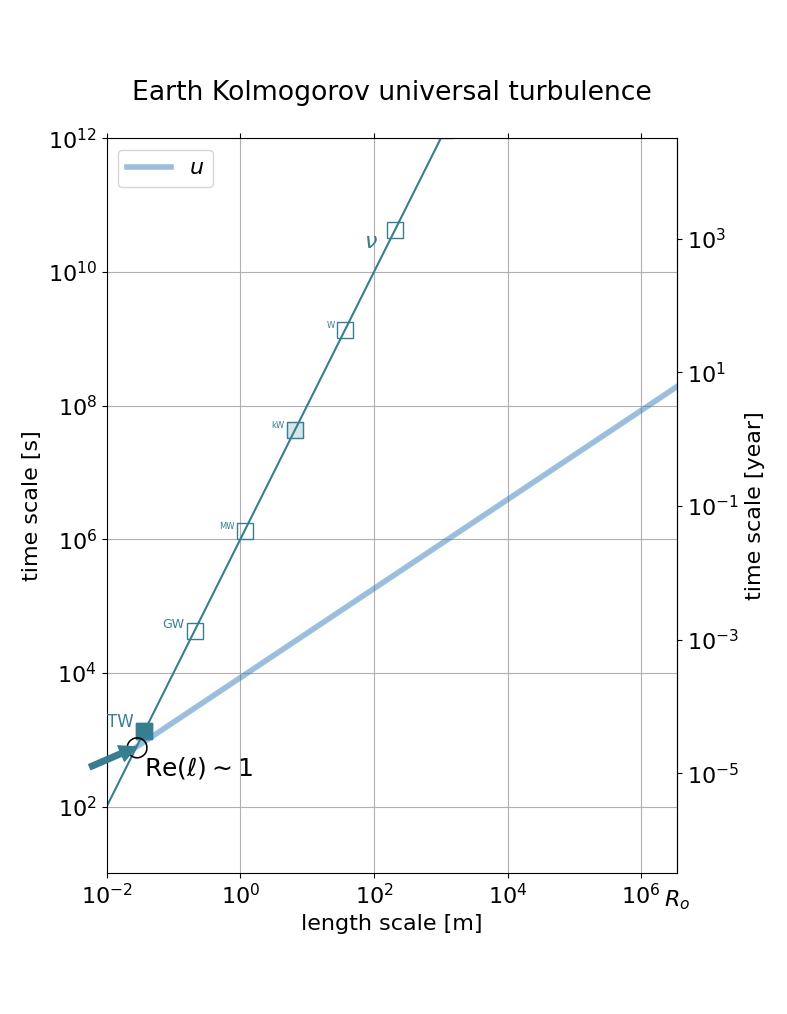}
		\caption{\tauell regime diagram for Kolmogorov's universal turbulence \citep{kolmogorov1941}.
		Teal line labeled $\nu$ is viscous dissipation line $\tau_\nu(\ell)=\ell^2/\nu$.
		The thick blue line is eddy turnover time $\tau_u(\ell)=\epsilon^{-1/3} \ell^{2/3}$ inferred from Kolmogorov's law, assuming that energy is injected at core radius length-scale $\ell=R_o$.
		These two lines intersect where the $\ell$-scale Reynolds number $\Re(\ell) \sim 1$, as marked by a circle.
		Total energy dissipation rate $\mathcal{P}_{diss}$ can be read at this intersection (blue arrow), using square markers drawn and labelled along the viscous line. Markers are a factor of $10^3$ apart, the 1 TW marker being filled.
		}
		\label{fig:K41}
	\end{figure}
\end{centering}

We first illustrate the construction rules of \tauell diagrams with the simple example of Kolmogorov's universal turbulence \citep{kolmogorov1941}.
Although this is not the kind of turbulence we expect in planetary cores, we pick a range of length scales and timescales typical of the Earth core.
Figure \ref{fig:K41} is a log-log plot of timescales spanning a range from $10$ s to $32 \, 000$ years versus length-scales from 1 cm to $R_o=3480$ km, the radius of the core.

\subsubsection{$\tau_\nu(\ell)$ line}
We pick a viscosity value $\nu = 10^{-6}$ m$^2$ s$^{-1}$ and draw the  $\tau_\nu(\ell)$ viscous dissipation line:
\begin{equation}
	\tau_\nu(\ell) = \ell^2/\nu.
		\label{eq:nu}
\end{equation}

\subsubsection{$\tau_u(\ell)$ line}
In Kolmogorov's universal turbulence, kinetic energy cascades down from large length-scales to small scales, from the energy injection scale down to the viscous dissipation scale.
The range in between is called the inertial range.
The kinetic energy density spectrum $E(k)$ in the inertial range obeys Kolmogorov's law:
\begin{equation}
	E(k) = C_K \epsilon^{2/3} k^{-5/3},
	\label{eq:K41}
\end{equation}
where $k$ is the wavenumber, $\epsilon$ is the energy injection (and dissipation) rate per unit mass, and $C_K$ is Kolmogorov's dimensionless constant, of order 1.

To build line $\tau_u(\ell)$, we need to convert kinetic energy density into velocity.
It is common to define an `eddy turnover time' as $\tau_u(\ell) = \ell/u(\ell)$, with $u^2(\ell) \sim E(k) k$ and $\ell \sim 1/k$ (see Appendix \ref{sec:conversion}).
This translates into:
\begin{equation}
	\tau_u(\ell) \simeq \ell^{3/2} \left[ E(\ell^{-1}) \right]^{-1/2}.
	\label{eq:taufromE}
\end{equation}
Dropping prefactor $C_K$, Kolmogorov's law yields:
\begin{equation}
	\tau_u(\ell) \simeq  \epsilon^{-1/3} \ell^{2/3} .
	\label{eq:tauK41}
\end{equation}

We draw this $\tau_u(\ell)$ line in Figure \ref{fig:K41}, assuming that the energy injection length-scale is $R_o$, and choosing an injection timescale, which will be discussed later.

\subsubsection{$\ell$-scale Reynolds number}
We terminate line $\tau_u(\ell)$ where it hits viscous line $\tau_\nu(\ell)$.
This intersection yields Kolmogorov microscales $(\ell_K,\tau_K)$, for which $\tau_K \equiv \tau_u(\ell_K) = \ell_K^2/\nu$, \textit{i.e.}, $u(\ell_K)\ell_K / \nu = 1$.
Defining an $\ell$-scale Reynolds number $\Re(\ell) = u(\ell) \ell / \nu$, we note that the intersection of the eddy turnover time line $\tau_u(\ell)$ with the viscous line $\tau_\nu(\ell)$ occurs at $\Re(\ell)=1$. It marks the transition from the inertial cascade at large scale to the viscous dissipation regime at small scale.

\subsubsection{Power dissipation markers}
In Kolmogorov's theory, the energy injected at large scale cascades down with no loss to small scales at which viscous dissipation takes place.
This dissipation range starts at the intersection of lines $\tau_u$ and $\tau_\nu$, where $Re(\ell) \sim 1$, which defines Kolmogorov microscales $(\ell_K,\tau_K)$.
From equations (\ref{eq:nu}) and (\ref{eq:tauK41}), we deduce:
\begin{equation}
	\epsilon = \frac{\ell_K^2}{\tau_K^3} = \frac{\nu}{\tau_K^2} = \frac{u^2(\ell_K)}{\tau_K},
	\label{eq:epsilon}
\end{equation}
where the last expression shows that dissipation rate per unit mass $\epsilon$ equals kinetic energy at the microscale divided by eddy turnover time at that scale.
We also see that we can attribute to each $\tau_\nu$ value a dissipation rate per unit mass.
Multiplying by the mass of the system, we obtain the total dissipated power.
In Figure \ref{fig:K41}, we thus draw power markers along the viscous dissipation line, using the mass of the outer core $M_o = 1.835 \; 10 ^{24}$ kg.
Markers are a factor of one thousand apart and are labeled.
In Figure \ref{fig:K41}, the timescale for energy injection at length-scale $R_o$ was chosen to yield a dissipated power $\mathcal{P}_{diss} = 3$ TW, marked by the blue arrow. 

\subsubsection{Energy}
In Kolmogorov's universal turbulence, kinetic energy is dominated by large length-scales.
In \tauell diagrams, we retrieve kinetic energy $\mathcal{E}_k$ from the square of the inverse of $\tau_u(R_o)$ since:
\begin{equation}
    	\mathcal{E}_k = \frac{1}{2} \int_{V_o} {\rho [u(\mathbf{r})]^2 dV} \sim M_o [u(R_o)]^2 = \frac{M_o R_o^2} {\tau_u^2(R_o)},
	\label{eq:E}
\end{equation}
where $V_o$ is the volume of the liquid core.
We use this property to compare the amplitudes of the various energy reservoirs when dealing with planetary cores.
Note that this applies as long as the slope of the $E(k)$ kinetic energy density spectrum is negative (\textit{i.e.}, slope of $\tau_u(\ell)$ is less than $3/2$), in order for large-scale energy to dominate.\\

\noindent
\fbox{%
\begin{minipage}{0.95\columnwidth}
    \textbf{Key properties of \tauell regime diagrams}\\
 
    \begin{itemize}
    \item \tauell regime diagrams gather in a simple graphical representation many of the ingredients that control the dynamics of a turbulent fluid system.
    
    \item In \tauell regime diagrams, intersections of lines $\tau_x(\ell)$ and $\tau_y(\ell)$ of physical phenomena $x$ and $y$ occur where $\ell$-scale dimensionless number $Z(\ell) = \frac{\tau_y(\ell)}{\tau_x(\ell)}$ equals 1.
    They mark a change in the system's dynamical regime.
    
    \item Usual integral-scale values of dimensionless numbers are obtained from the ratio of relevant $\tau_x(\ell)$ and $\tau_y(\ell)$ times at integral scale $\ell=R_o$.

    \item Total dissipated power can be marked along $\tau(\ell)$ lines of dissipative phenomena.
    
    \item Energies of different types (kinetic, gravitational, magnetic) are represented by the inverse square of corresponding $\tau(R_o)$.

    \item \tauell regime diagrams are a useful tool to infer or test turbulence scenarios.
    They are \emph{not} a theory of turbulence.
	
     \end{itemize}
	
\end{minipage}
}

\section{Physical phenomena in planetary cores and dimensionless numbers }
\label{sec:phenomena}

We now turn our attention to planetary cores.
Flow within planetary cores are mostly powered by thermal or thermo-compositional convection.
They often produce a magnetic field.
Most importantly, these flows occur in a rotating spherical system.

In this section, we introduce the \tauell lines these physical phenomena contribute.
We illustrate the resulting \tauell regime diagram template, using values pertaining to Earth's core, and relate the diagram to various dimensionless numbers used to characterize planetary core dynamics.

\subsection{Physical phenomena and their \tauell expressions}
\label{sec:scales}

Table \ref{tab:scales} gives the expressions of major $\tau(\ell)$ times pertaining to planetary cores.
$\tau_\nu(\ell)$ and $\tau_u(\ell)$ times have been introduced in section \ref{sec:K41}.
Convection adds thermal diffusion and buoyancy scales.
Rotation, spherical boundaries, and magnetic field contribute key additional timescales.

\begin{centering}
	\begin{figure}
		\begin{subfigure}[b]{=7.7cm}
			\includegraphics[width=\textwidth]{ 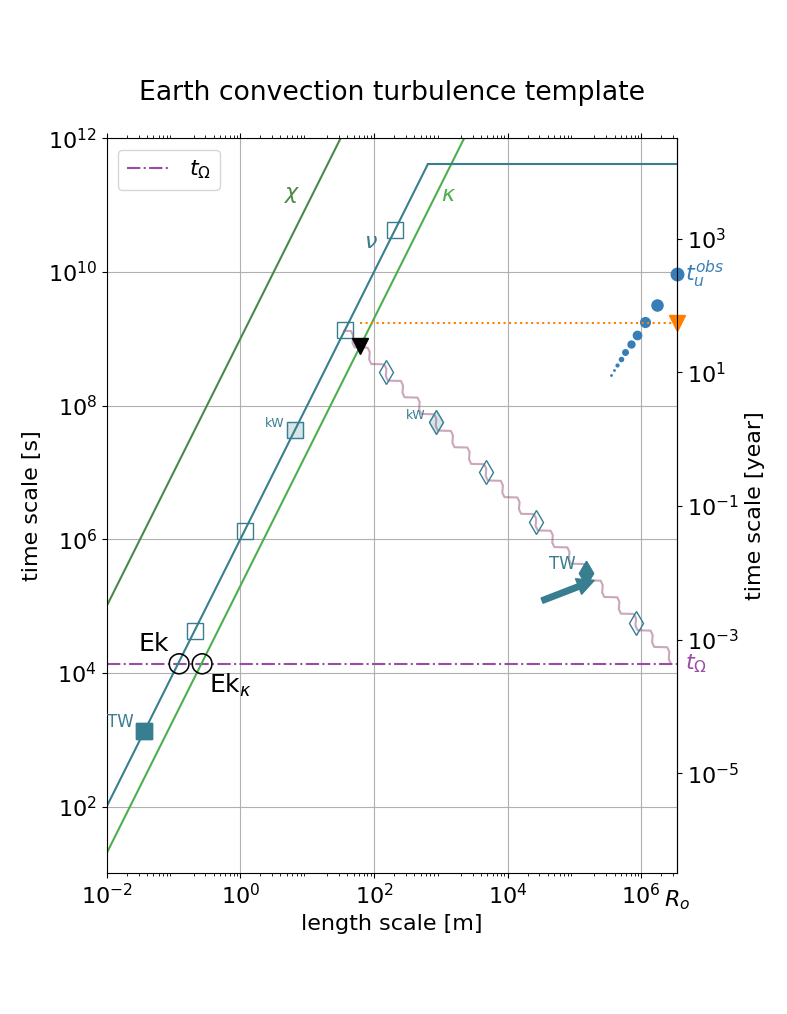}
			\caption{Template for Earth's core rotating convection.}
			\label{fig:convection_template}
		\end{subfigure}
		\begin{subfigure}[b]{=7.7cm}
			\includegraphics[width=\textwidth]{ 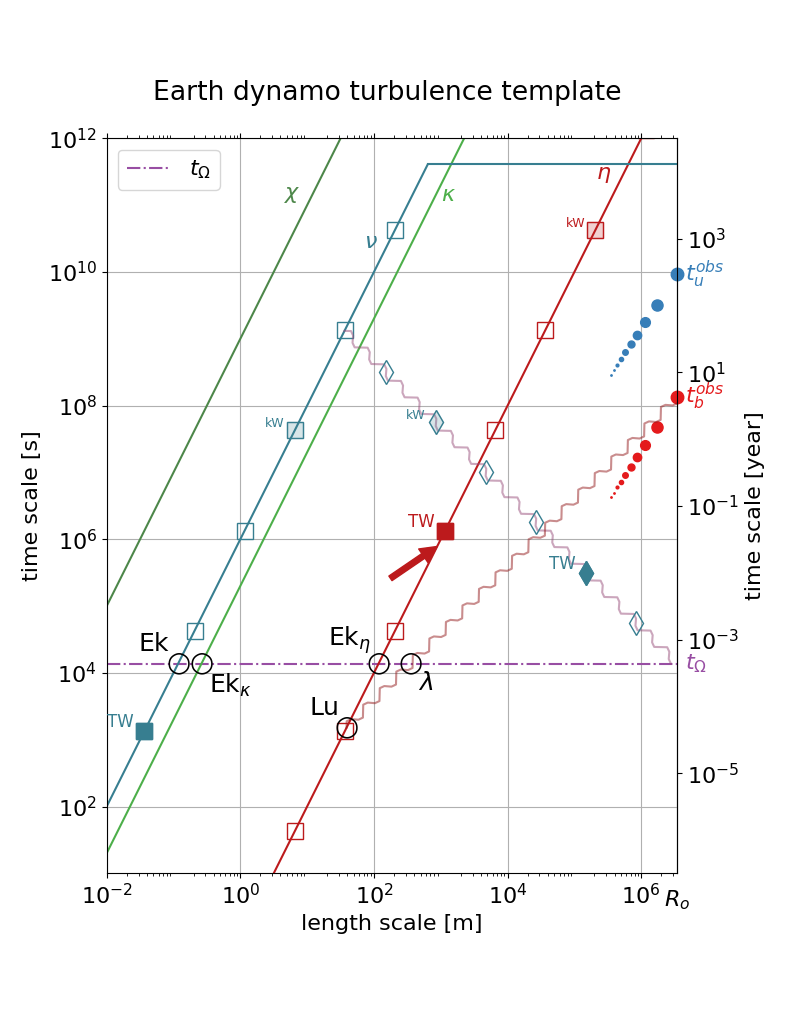}
			\caption{Template for Earth's core rotating convective dynamo.}
			\label{fig:dynamo_template}
		\end{subfigure}
		\caption{\tauell regime diagram templates for the Earth's core.
		The time-length relationships of relevant physical phenomena, as given in Table \ref{tab:scales}, are drawn in a log-log plot of timescale versus length-scale, using Earth core values from Table \ref{tab:properties} presented in section \ref{sec:core}.
		(a) \emph{Template ignoring the magnetic field.}
		The steep solid lines labeled $\chi$, $\nu$ and $\kappa$ are diffusion times $\tau_\chi(\ell)$, $\tau_\nu(\ell)$, and $\tau_\kappa(\ell)$, respectively.
		The dash-dot horizontal line is the rotation time $t_\Omega$.
		The Rossby line $\tau_{Rossby}(\ell)$ is drawn as a wavy line pinned to $t_\Omega$.
		Markers along lines $\tau_\nu(\ell)$ and $\tau_{Rossby}(\ell)$ indicate viscous power dissipation.
		Markers are a factor of 1000 apart, the 1 TW marker being filled.
		The circle labeled $\Ek$ at the intersection of the viscous $\tau_\nu(\ell)$ and $t_\Omega$ lines mark the length-scale at which the $\ell$-scale Ekman number equals 1.
		Same thing for the thermal Ekman number $\Ek_\kappa(\ell)$.	
		(b) \emph{Template including the magnetic field.}
		Same as (a) with additional lines and labels brought by the magnetic field.
		The steep solid red line labeled $\eta$ is the magnetic diffusion line $\tau_\eta(\ell)$.
		Markers along that line indicate Ohmic power dissipation.
		Markers are a factor of 1000 apart, the 1 TW marker being filled.
		The Alfv\'en line $\tau_{Alfven}(\ell)$ is drawn as a wavy line pinned to the observed large-scale magnetic field Alfv\'en time $t_b^{obs}$.
		Circles labeled $\Ek_\eta$, $\Lu$ and $\lambda$ at line intersections mark scales at which the corresponding $\ell$-scale dimensionless number (see Table \ref{tab:dimensionless}) equals 1.
		See sections \ref{sec:onset_note} and \ref{sec:core} for more information.
		}
		\label{fig:template}
	\end{figure}
\end{centering}

We discuss the origin and meaning of these various $\tau(\ell)$ scales, and illustrate the \tauell template they provide in Earth's core example (Figure \ref{fig:template}), using its properties presented in section \ref{sec:core}.
We first ignore the magnetic field and build the template of Figure \ref{fig:convection_template}.

\begin{table}
	\begin{center}
		\begin{tabular}{lcl}
			time & expression & phenomenon \\
			\hline
			$\tau_\nu(\ell)$ & $\ell^2/\nu$ & viscous dissipation\\
			$\tau_\kappa(\ell)$ & $\ell^2/\kappa$ & thermal diffusion\\
			$\tau_\chi(\ell)$ & $\ell^2/\chi$ & compositional diffusion\\
			$\tau_\eta(\ell)$ & $\ell^2/\eta$ & magnetic dissipation\\
			$t_\Omega$ & $1/\Omega$ & rotation\\
			$\tau_{Rossby}(\ell)$ & $R_o/\Omega \ell$ & Rossby wave propagation\\
			$\tau_{Alfven}(\ell)$ & $\ell \sqrt{\rho\mu}/B_0$ & Alfv\'en wave propagation \\
			& & \\
			$\tau_\rho(\ell)$ & $\sqrt{\frac{\ell}{g}\frac{\rho}{|\Delta\rho(\ell)|}}$ & buoyancy (or free-fall)\\
			$\tau_u(\ell)$ & $\ell / u(\ell)$ & eddy turnover\\
			$\tau_b(\ell)$ & $\ell \sqrt{\rho\mu}/b(\ell)$ & Alfv\'en wave collision\\
			\hline
			\smallskip
		\end{tabular}
		\caption{\label{tab:scales} Notation and expression of $\tau(\ell)$ times of relevant physical phenomena for planetary cores.
\emph{Fluid properties:} density $\rho$; kinematic viscosity $\nu$; thermal and compositional diffusivities $\kappa$ and $\chi$, respectively; magnetic diffusivity $\eta$;  magnetic permeability $\mu$.
\emph{System properties:} radius $R_o$; gravity $g$; rotation rate $\Omega$; large-scale magnetic field $B_0$.
\emph{Turbulent flow properties:}
$\Delta \rho(\ell)$, $u(\ell)$ and $b(\ell)$ are $\ell$-scale density anomaly, flow velocity, and magnetic field intensity, respectively.
We also write $\tau_\eta(R_o)$ as $T_\eta$ for short.
Adapted from Table 1 of Chapter 8.06 of \textit{Treatise on Geophysics} \citep{nataf2015} with permission.}
	\end{center}
\end{table}

\subsubsection{diffusion}
Lines labeled  $\nu$, $\kappa$ and  $\chi$ are diffusion \tauell lines. Timescales of diffusive phenomena all share the same form $\tau(\ell) = \ell^2/D$, where diffusivity $D$ is $\nu$, $\kappa$ or  $\chi$ depending upon which field diffuses: momentum, temperature, or composition, respectively.

\subsubsection{convection}
We introduce a `buoyancy' or `free-fall' timescale $\tau_\rho(\ell) = \sqrt{\frac{\ell}{g} \frac{\rho}{|\Delta\rho(\ell)|}}$, which is the time it takes for a parcel of fluid with density anomaly $\Delta\rho(\ell)$ to rise or sink a distance $\ell$ in the absence of diffusion.
$\tau_\rho(\ell)$ relates to density anomaly $\Delta\rho$ at length-scale $\ell$.
Density anomaly, flow velocity and magnetic field constitute the three fields for which we seek an adequate turbulent description.

We have shown in section \ref{sec:K41} that the value of $\tau_u$ at integral scale $R_o$ measures the kinetic energy of the flow.
Similarly, gravitational energy $\mathcal{E}_g$ is measured by $\tau_\rho(R_o)$ (as long as the slope of $\tau_\rho(\ell)$ is less than $3/2$) since:
\begin{equation}
    	\mathcal{E}_g = \frac{1}{2} \int_{V_o} { g r \Delta\rho(\mathbf{r}) dV} \sim M_o g R_o \frac{\Delta\rho(R_o)}{\rho} = \frac{M_o R_o^2} {\tau_\rho^2(R_o)}.
	\label{eq:E_g}
	\end{equation}

\subsubsection{rotation}
Rotation is a crucial ingredient of planetary core dynamics.
It adds one important time in our \tauell regime diagram: the rotation time $t_\Omega = \Omega^{-1}$, \textit{i.e.,} one day divided by $2 \pi$.
Physical phenomena operating at timescales smaller than $t_\Omega$ are not influenced by planet's spin, while those with longer timescales feel the effect of rotation.
We thus draw a horizontal line at $t_\Omega$ in the diagram of Figure \ref{fig:convection_template}.
The intersection of this line with the viscous line yields Ekman layer's thickness $\ell_E = \sqrt{\nu/\Omega}$.
Viscous forces are balanced by Coriolis acceleration in these thin Ekman layers.
$\ell_E$ is the only length-scale one can build from $\nu$ and $\Omega$ alone, and it controls friction, hence viscous dissipation, that takes place at boundaries.
It turns out that boundaries bring up new important dynamical constraints and scales.

\subsubsection{rotation and spherical boundaries}
\label{sec:QG}
We will not review here the vast literature on rotating fluids in containers.
The book of \citet{greenspan1968} remains amazingly central.
At this stage, let us simply recall that Navier-Stokes equation reduces to \emph{geostrophic equilibrium} when Coriolis acceleration dominates:
\begin{equation}
	2 \rho \mathbf{\Omega} \times \mathbf{u} = -\nabla P,
	\label{eq:geostrophic}
\end{equation}
where $\mathbf{u}$ and $P$ are fluid velocity and pressure, respectively.
Taking the curl of this equation yields \emph{Proudman-Taylor constraint}:
\begin{equation}
	\frac{\partial \mathbf{u}}{\partial z} = 0,
	\label{eq:PT}
\end{equation}
where $z$ coordinate is parallel to vector $\mathbf{\Omega}$.
Reintroducing acceleration term $\rho \partial_t \mathbf{u}$ allows for the propagation of \emph{inertial waves}.

Proudman-Taylor constraint would inhibit all fluid motions in a rotating fluid bounded by a solid container.
Accounting for the presence of a thin Ekman layer that accommodates a velocity jump between the fluid bulk and the boundary, \emph{geostrophic} flows are allowed, which follow contours of equal fluid column-height (measured in the $z$-direction), \textit{i.e.,} azimuthal flows in a spherical container.
In most situations, \emph{Quasi-geostrophic} (QG) fluid motions are also observed, which approximately satisfy Proudman-Taylor constraint (\textit{i.e.,} $z$-invariance) in the bulk (at least for one component, typically the azimuthal velocity).

It is important to note that Proudman-Taylor constraint is established by the propagation of inertial waves in the fluid, and is effective only when they had time to reach a boundary.
Thus, a localized eddy of radius $\ell$ grows into a columnar vortex at a speed equal to $\Omega \ell$ \citep{davidson2006}.
This means that large eddies rapidly form quasi-geostrophic columns, while it takes more time for small eddies to form core-size columns.
Time for reaching quasi-geostrophy is thus given by:
\begin{equation}
	\tau_{Rossby}(\ell) = \frac{R_o}{\Omega \ell},
	\label{eq:Rossby}
\end{equation}
as written in Table \ref{tab:scales}.
This line is drawn as a wavy line in Figure \ref{fig:convection_template}.
It is pinned to time $t_\Omega$ at $\ell=R_o$, and we extend it until it reaches viscous line $\tau_\nu(\ell)$.

Note that $\tau_{Rossby}(\ell)$ also equals the time it takes for a Rossby wave of wavelength $\ell$ to propagate one wavelength (hence its name) \citep{nataf2015}.
The intersection of the $\tau_u(\ell)$ line with the Rossby line has $\ell =\sqrt{u(\ell) R_o/ \Omega}$, which defines a \emph{Rhines scale} (originally more precisely defined as $\ell =\sqrt{u/\beta}$ in a thin shell, where $\beta = 2 \Omega \sin\theta/R_o$ is the northward gradient of Coriolis frequency at colatitude $\theta$ \citep{rhines1975}, here extended to a wide gap \citep{busse1970,schopp1997}).

Flow is quasi-geostrophic for scales above line $\tau_{Rossby}(\ell)$.
In the triangle formed by the Rossby line, the viscous line and line $t_\Omega$, flow structures are elongated parallel to the spin axis but not enough to reach both boundaries.
Flow is 3D beneath line $t_\Omega$.

\subsubsection{Quasi-geostrophic dissipation}
Quasi-geostrophic vortices dissipate kinetic energy by Ekman friction at no-slip boundaries of the liquid core.
We approximate energy loss rate $p_\ell$ of a single QG vortex of radius $\ell$ by:
\begin{equation}
	p_\ell = \rho \nu \frac{u^2(\ell)}{\ell_E^2} \ell_E \ell^2,
\end{equation}
with $\ell_E$ the Ekman layer thickness.
Summing contributions of all $\ell$-scale QG vortices filling the entire core, we obtain the total power dissipated by Ekman friction $\mathcal{P}_{QG}$.
Dividing by the mass of the core, we obtain the QG viscous dissipation rate per unit mass $\epsilon_{QG}$ at scale $\ell$ as:
\begin{equation}
	\epsilon_{QG} = \nu \frac{u^2(\ell)}{R_o \ell_E}
	\label{eq:epsilonQG}
\end{equation}
Ekman friction matters for QG vortices.
We therefore draw the corresponding power markers on the Rossby line, above which flow is quasi-geostrophic.
Time $\tau_{QG}$ for which total viscous dissipation by Ekman friction equals $\mathcal{P}_{QG}$ is then obtained from equation (\ref{eq:epsilonQG}) as:
\begin{equation}
	\tau_{QG} = \left[ \frac{M_o R_o}{\mathcal{P}_{QG}} \sqrt{\frac{\nu}{\Omega^3}} \right]^{1/4}.
	\label{eq:tauQG}
\end{equation}
This provides markers drawn in Figure \ref{fig:convection_template} as diamonds along the Rossby line, a factor of one thousand apart, the TW marker being filled.

Going back to the viscous line $\tau_\nu(\ell)$, we terminate it at the spin-up time $R_o/\sqrt{\nu \Omega} \sim 13 \times 10^3$ years, which is the time it takes for a change in the outer boundary spinning rate to be transmitted to the entire volume of the core.

\subsubsection{A note on convection onset}
\label{sec:onset_note}
We can already illustrate an interesting insight provided by this simple \tauell template, by adding the scales appearing at the onset of convection in this rotating spherical system.
This topic has a long history, starting with the pioneer studies of \citet{roberts1968} and \citet{busse1970}, followed by \citet{jones2000, dormy2004, zhang2007}.
It is found that convection sets in as a travelling thermal Rossby wave, forming columns aligned with the spin axis, whose width is controlled by viscosity in the bulk of the spherical shell.
A \tauell translation of the length-scale, period, and Rayleigh number at convection onset is given in Appendix \ref{sec:onset}.

In Figure \ref{fig:convection_template}, a black triangle marks the length-scale and period at convection onset.
It lies at the intersection of the Rossby and thermal diffusion lines, close to the viscous diffusion line, as expected for a viscously-controlled quasi-geostrophic thermal Rossby wave.
We read a column-width of about 100 m.
Is this the typical convective length-scale in the Earth's core?
We will get back to this question in section \ref{sec:length-scale}.
The orange triangle at $\ell=R_o$ marks the critical free-fall time $T_\rho^c$ deduced from the critical Rayleigh number (see Appendix \ref{sec:onset}).

\subsubsection{magnetic field, magnetic dissipation, magnetic energy}
Magnetic fields are often produced and sustained by dynamo action within planetary cores.
We now add the magnetic field to build the template of Figure \ref{fig:dynamo_template}. 
The red line labeled $\eta$ is the magnetic diffusion line $\tau_\eta(\ell)$.
Magnetic dissipation markers are labeled along that line, following the same rule as in equation (\ref{eq:epsilon}).

The presence of a magnetic field allows the propagation of magnetohydrodynamic waves called Alfv\'en waves \citep{alfven1942}.
In a uniform magnetic field $B_0$, these waves propagate at speed $V_A = B_0 / \sqrt{\rho \mu}$, where $\mu$ is fluid's magnetic permeability.
Assuming a large-scale magnetic field $B_0$, we construct line $\tau_{Alfven}(\ell) = \ell \sqrt{\rho \mu}/B_0$, the time it takes for an Alfv\'en wave to propagate over a distance $\ell$.
It is drawn as a red wavy line in Figure \ref{fig:dynamo_template}.

To describe the magnetic field in the system, we define a similar timescale, replacing $B_0$ by $\ell$-scale magnetic field $b(\ell)$.
Note that $\tau_b(R_o) = \tau_{Alfven}(R_o)$ provides the magnitude of magnetic energy $\mathcal{E}_m$ (as long as the slope of $\tau_b(\ell)$ is less than $3/2$), since:
\begin{equation}
    	\mathcal{E}_m = \frac{1}{2 \mu} \int_{V_o} {[b(\mathbf{r})]^2 dV} \sim \frac{M_o}{\rho \mu} [b(R_o)]^2 = \frac{M_o R_o^2}{\tau_b^2(R_o)}.
	\label{eq:E_m}
\end{equation}

\subsection{Dimensionless numbers}
\label{sec:dimensionless}

In order to connect to the huge literature pertaining to geophysical and astrophysical fluid dynamics, it is important to relate our \tauell regime diagrams to widely used dimensionless numbers.
Dimensionless numbers provide the minimum number of parameters needed to describe a physical system.
They permit a comparison of widely different systems that yield the same dimensionless numbers.
These numbers are dimensionless combinations of properties and field variables that appear when the equations governing the dynamics of the system under study are made dimensionless by normalizing their various terms by `typical scales'.

For example, Reynolds number for a system of size $L$ will be written: $\Re = U L / \nu$, where $U$ is a typical fluid velocity, and $\nu$ kinematic viscosity.
Usually, it is when this dimensionless number is of order 1 that a change of regime occurs.
In this example: a change between a regime where momentum diffusion dominates over advection when $\Re<1$ to one where advection dominates for $\Re>1$.

Most dimensionless numbers can be written as the ratio of two times.
In our approach, we define length-scale dependent dimensionless numbers, constructed as the ratios of the timescales of the relevant physical phenomena at that length-scale.
We thus define $\ell$-scale Reynolds number as: $\Re(\ell) = \tau_\nu(\ell) / \tau_u(\ell)$, where $\tau_\nu(\ell)$ is momentum diffusion timescale at length-scale $\ell$, while $\tau_u(\ell)$ is the overturn time of a vortex of radius $\ell$.
Table \ref{tab:dimensionless} gives the expressions of $\ell$-scale dimensionless numbers pertaining to planetary liquid core dynamics.

\begin{table*}
 	\begin{center}
		\begin{tabular}{lccl}
			number & expression & time ratio & name \\
			\hline
			\smallskip
			$\Re(\ell)$ & {\Large $\frac{u(\ell)\ell}{\nu}$} & {\Large $\frac{\tau_\nu(\ell)}{\tau_u(\ell)}$} & Reynolds \\
			\smallskip
			$\Ra(\ell)$ & {\Large $\frac{g \ell^3 |\Delta\rho(\ell)|/\rho}{\kappa \nu}$} & {\Large $\frac{\tau_\kappa(\ell) \, \tau_\nu(\ell)}{\tau_\rho^2(\ell)}$} & Rayleigh \\
			\smallskip
			$\Ek(\ell)$ & {\Large $\frac{\nu}{\Omega \ell^2}$} & {\Large $\frac{t_\Omega}{\tau_\nu(\ell)}$} & Ekman \\
			\smallskip
			$\Ek_\kappa(\ell)$ & {\Large $\frac{\kappa}{\Omega \ell^2}$} & {\Large $\frac{t_\Omega}{\tau_\kappa(\ell)}$} & thermal Ekman \\
			\smallskip
			$\Ek_\eta(\ell)$ & {\Large $\frac{\eta}{\Omega \ell^2}$} & {\Large $\frac{t_\Omega}{\tau_\eta(\ell)}$} & magnetic Ekman \\
			\smallskip
			$\Ro(\ell)$ & {\Large $\frac{u(\ell)}{\Omega \ell}$} & {\Large $\frac{t_\Omega}{\tau_u(\ell)}$} & Rossby \\
			\smallskip
			$\Roff(\ell)$ & {\Large $\frac{\sqrt{g \ell |\Delta \rho(\ell)|/\rho}}{\Omega \ell}$} & {\Large $\frac{t_\Omega}{\tau_\rho(\ell)}$} & free-fall Rossby \\
			\smallskip
			$\Rm(\ell)$ & {\Large $\frac{u(\ell)\ell}{\eta}$} & {\Large $\frac{\tau_\eta(\ell)}{\tau_u(\ell)}$} & magnetic Reynolds \\
			\smallskip
			$\Lu(\ell)$ & {\Large $\frac{\ell B_0}{\eta \sqrt{\rho \mu}}$} & {\Large $\frac{\tau_\eta(\ell)}{\tau_{Alfven}(\ell)}$} & Lundquist \\
			\smallskip
			$\Lambda_d(\ell)$ & {\Large $\frac{b(\ell) B_0}{\rho \mu u(\ell) \Omega \ell}$} & {\Large $\frac{\tau_u(\ell) \, t_\Omega}{\tau_b(\ell) \, \tau_{Alfven}(\ell)}$} & dynamical Elsasser \\
			\smallskip
			$\lambda(\ell)$ & {\Large $\frac{B_0}{\sqrt{\rho \mu} \Omega \ell}$} & {\Large $\frac{t_\Omega}{\tau_{Alfven}(\ell)}$} & Lehnert \\
			\hline
		\end{tabular}
 	\end{center}			
	\caption{Expressions of $\ell$-scale dimensionless numbers.
These numbers are also expressed as ratios of characteristic $\ell$-scale times, which are defined in Table \ref{tab:scales}. One recovers the classical expression of these numbers at integral scale by setting $\ell = R_o$.
Adapted from Table 2 of Chapter 8.06 of \textit{Treatise on Geophysics} \citep{nataf2015} with permission.
}
	\label{tab:dimensionless}
\end{table*}

In \tauell regime diagrams, the intersection of the $\tau_x(\ell)$ and $\tau_y(\ell)$ lines of physical phenomena $x$ and $y$ occurs where $\ell$-scale dimensionless number $Z(\ell) = \tau_y(\ell) / \tau_x(\ell)$ equals 1.
Each such intersection marks a change in the system's dynamic regime.

In Figures \ref{fig:convection_template} and \ref{fig:dynamo_template}, we have labeled several line intersections, where specific dimensionless numbers equal 1.
The intersections of line $t_\Omega$ and the $\nu$, $\kappa$ and $\eta$ lines indicate where corresponding $\ell$-scale Ekman numbers equal 1 in the \tauell plane.
The intersection of the Alfv\'en and magnetic diffusion lines defines where the $\ell$-scale Lundquist number equals 1, marking a change from propagating Alfv\'en waves at larger scales to damped waves at smaller scales.
Similarly, the intersection of the Alfv\'en and $t_\Omega$ lines yields $\lambda(\ell) \sim 1$, where $\lambda$ is the Lehnert number \citep{lehnert1954,jault2008}.
System rotation favors quasi-geostrophic Alfv\'en waves at timescales above this intersection.

More dimensionless numbers, such as $\Re$, $\Rm$, $\Ra$, $\Ro$, $\Roff$, and $\Lambda_d$, will appear when we plot lines $\tau_\rho(\ell)$, $\tau_u(\ell)$ and $\tau_b(\ell)$ of the system's density, velocity and magnetic fields for the different turbulence scenarios we will explore.

\subsection{A word on Earth's core properties}
\label{sec:core}

\begin{table*}
 	\begin{center}
		\begin{tabular}{lcll}
			symbol & value & unit & property \\
			\hline
			$\nu$           & $10^{-6}$               & m$^2$ s$^{-1}$ & kinematic viscosity \\
			$\kappa$     & $5 \times 10^{-6}$         & m$^2$ s$^{-1}$ & thermal diffusivity \\
			$\chi$          & $10^{-9}$               & m$^2$ s$^{-1}$ & compositional diffusivity \\
			$\eta$          & $1$                        & m$^2$ s$^{-1}$ & magnetic diffusivity \\
			$\rho$          & $10.9 \times 10^3$        & kg m$^{-3}$ & density \\
			$\alpha$       & $1.2 \times 10^{-5}$      & K$^{-1}$ & thermal expansion coefficient \\
			$C_P$          & 850                         & J kg$^{-1}$ K$^{-1}$ & specific heat capacity \\
			$R_o$           & $3.48 \times 10^{6}$      & m & core radius \\
			$R_i$             & $1.22 \times 10^{6}$     & m & inner core radius \\
			$M_o$            & $1.835 \times 10^{24}$ & kg & outer core mass \\
			$g$                 & $8$                        & m s$^{-2}$ & gravity \\
			$t_{\Omega}$ & $1.38 \times 10^{4}$     & s & Earth's rotation time (\textit{i.e.} $1/2\pi$ day) \\
			$\mathcal{P}_{diss}$ & $3 \times 10^{12}$ & W & available convective power \\
			& & & \\
			$t_u^{obs}$ & $9 \times 10^{9}$ & s & $R_o$-scale core flow time (\textit{i.e.} $\simeq 300$ years) \\
			$t_b^{obs}$ & $1.4 \times 10^{8}$ & s & $R_o$-scale Alfv\'en wave time (\textit{i.e.} $\simeq 4$ years) \\
			\hline
		\end{tabular}
 	\end{center}
	\caption{Properties of Earth's core. See \citet{olson2015} for details and uncertainties.
Adapted from Table 3 of Chapter 8.06 of \textit{Treatise on Geophysics} \citep{nataf2015} with permission.
}
	\label{tab:properties}
\end{table*}

Core properties used to build the \tauell diagrams of Figure \ref{fig:template} are listed in Table \ref{tab:properties}.
Most are taken from Peter Olson's review in  \textit{Treatise on Geophysics} \citep{olson2015}.
Some of them are known with great precision (to about 1\textperthousand \, for core radius $R_o$ and liquid core mass $M_o$), but others are poorly constrained (to about 1 or 2 orders of magnitude for viscosity $\nu$ and compositional diffusivity $\chi$).
In addition, most physical properties are expected to vary with radius.
None of these (important) subtleties are taken into account in the `fuzzy' approach we advocate for building \tauell regime diagrams.
Note that we systematically drop all numerical prefactors, including $2 \pi$.

As noted in section \ref{sec:introduction}, the power available to drive the dynamics of the system under study is a key ingredient.
It largely controls the different turbulence regimes the system will experience.
Thermal evolution of the Earth has received considerable attention (see \citet{nimmo2015, landeau2022,driscoll2023} for reviews).
It is now well established that the dynamics of Earth's core today is powered by its slow cooling, enhanced by the resulting growth of the solid inner core.
As iron-nickel alloy crystallizes at its surface, it releases latent heat and light elements that drive convection and power the geodynamo.

Despite uncertainties on isentropic heat flux, the available convective power is found to be in the range $0.8-5 \times 10^{12}$ W for present-day core \citep{nimmo2015,landeau2022}.
We adopt value $\mathcal{P}_{diss} = 3$ TW.
This value is pointed by a blue arrow on the Rossby line in Figure \ref{fig:convection_template}, and by a red arrow on the magnetic dissipation line in Figure \ref{fig:dynamo_template}.

The last two rows in Table \ref{tab:properties} are not used to build \tauell diagrams, but instead to test their relevance.
Large-scale vortex turnover time $t_u^{obs} \simeq 300$ years is retrieved from core flow inversions of magnetic secular variation \citep[\textit{e.g.,}][]{pais2008}.
Large-scale Alfv\'en wave propagation time $t_b^{obs} \simeq 4$ years is deduced from the discovery and analysis of `torsional oscillations' in Earth's core \citep{gillet2010}.

It is also observed that the Lowes-Mauersberger spectrum of magnetic energy is flat at the core-mantle boundary up to harmonic degree 10 \citep[\textit{e.g.},][]{langlais2014}.
This means that energy is independent of length-scale in this scale-range, which translates into a $\tau_b(\ell) \propto \ell^{3/2}$ trend at large-scale (see Appendix \ref{sec:spectra_LM}).
Similarly, core flow inversions favor an almost flat harmonic spectrum of kinetic energy up to degree 10 \citep{roberts2013,aubert2013,gillet2015,baerenzung2016}.
These trends are sketched by colored disks labeled $t_u^{obs}$ and $t_b^{obs}$ in Figure \ref{fig:template}.

\section{\tauell regime diagrams for a non-magnetic rotating convective core}
\label{sec:non-magnetic}
Building a turbulence scenario for a given system, starting from the power it dissipates, requires estimating the balance of forces at different length-scales.
We thus start by expressing the \tauell translation of expected relevant force balances.
We then focus on a quasi-geostrophic regime, which we illustrate with the \tauell diagram of an actual numerical simulation.
We then propose an idealized scenario of turbulent convection in Earth's core in the absence of a magnetic field.

\subsection{\tauell expression of force balances in rotating convection}
\label{sec:force_balance}
Let us start from the Navier-Stokes equation for deviations from hydrostatic equilibrium in an incompressible fluid under the Boussinesq approximation:
\begin{equation}
	\rho \left( \partial_t \mathbf{u} +  \mathbf{u} \cdot \nabla \mathbf{u} + 2 \bm{\Omega} \times \mathbf{u} \right) = -\nabla P + \Delta \rho \mathbf{g} + \rho \nu \nabla^2 \mathbf{u},
	\label{eq:NS}
\end{equation}
where the symbols have their usual meaning.
The acceleration term on the left-hand side includes advection and Coriolis, while the right-hand side figures pressure gradient, buoyancy, and viscous forces.

We already recalled in section \ref{sec:QG} that equation (\ref{eq:NS}) reduces to geostrophic equilibrium (\ref{eq:geostrophic}) when Coriolis acceleration strongly dominates, yielding Proudman-Taylor constraint (\ref{eq:PT}).
However, this equation is a diagnostic equation that can't be used on its own, as gets clear when equation (\ref{eq:NS}) is curled to obtain the vorticity equation (see \citet{jones2015} for a more complete treatment).
Reintroducing other accelerations and forces in the vorticity equation can lead to two different situations: (i) Proudman-Taylor constraint is broken and we get a three-term balance between Coriolis, inertia (or viscosity) and buoyancy; (ii) Coriolis acceleration is still dominant, and flow is quasi-geostrophic at leading order, with a small velocity gradient along the spin axis $(\bm{\Omega} \cdot \nabla) \mathbf{u}$, scaling as $1/\ell_\parallel$, where $\ell_\parallel \sim R_o$ \citep{julien2012}.

\subsubsection{Coriolis-Inertia-Archimedes (CIA)}
\label{sec:CIA}
Let's first consider the first situation, with a three-term balance of Coriolis, inertia and Archimedean forces.
At a given length-scale $\ell_\Omega$, retaining these forces in the curl of equation (\ref{eq:NS}) yields:
\begin{equation}
	\frac{\Omega u}{\ell_\Omega} \sim \frac{u^2}{\ell_\Omega^2} \sim \frac{\Delta \rho }{\rho} \frac{g}{\ell_\Omega}
	\label{eq:CIA}
\end{equation}
Translating in \tauell language, we get:
\begin{equation}
	t_\Omega \tau_u(\ell_\Omega) \sim \tau_u^2(\ell_\Omega) \sim \tau_\rho^2(\ell_\Omega),
	\label{eq:CIA_tauell}
\end{equation}
which implies:
\begin{equation}
	\tau_u(\ell_\Omega) \sim t_\Omega  \sim \tau_\rho(\ell_\Omega).
	\label{eq:CIA_tauell}
\end{equation}
This is the regime we expect when the $\tau_u(\ell)$ line reaches the $t_\Omega$ line, where $\Ro(\ell_\Omega) \sim 1$.

\subsubsection{Quasi-Geostrophic Coriolis-Inertia-Archimedes (QG-CIA)}
\label{sec:QG-CIA}
The second situation is more relevant for the Earth's core in the absence of a magnetic field.
We still consider a three-term balance of Coriolis, inertia and Archimedean forces, but in which the Coriolis term is reduced to its ageostrophic part:
\begin{equation}
	\frac{\Omega u}{\ell_\parallel} \sim \frac{u^2}{\ell_\perp^2} \sim \frac{\Delta \rho }{\rho} \frac{g}{\ell_\perp}
	\label{eq:QG-CIA}
\end{equation}
The first term corresponds to vortex stretching, the second one to vorticity advection, and the last one to vortex generation by buoyancy \citep{cardin1994,aubert2001,jones2015}.

Translating in \tauell language, and assuming $\ell_\parallel \sim R_o$, we get:
\begin{equation}
	\frac{R_o}{\ell_\perp}t_\Omega \tau_u(\ell_\perp) \sim \tau_\rho^2(\ell_\perp) \sim \tau_u^2(\ell_\perp).
	\label{eq:QG-CIA_tauell}
\end{equation}
Equation (\ref{eq:QG-CIA_tauell}) implies: 
\begin{equation}
	\tau_u(\ell_\perp) \simeq \frac{R_o}{\ell_\perp}t_\Omega \equiv \tau_{Rossby}(\ell_\perp),
\end{equation}
meaning that the $\tau_u(\ell)$ of a flow in QG-CIA balance plots on the Rossby line $\tau_{Rossby}(\ell)$ (see Table \ref{tab:scales}).
Note that where flow is quasi-geostrophic, the $\ell$ that we use to describe its dynamics in \tauell diagrams is the length-scale in the equatorial plane (\emph{i.e.,} the width of the columns).

\subsection{\tauell diagram of a remarkable numerical simulation}
\label{sec:Guervilly}
\begin{centering}
	\begin{figure}
		\begin{subfigure}[b]{6.8cm}	
			\includegraphics[width=\textwidth]{ 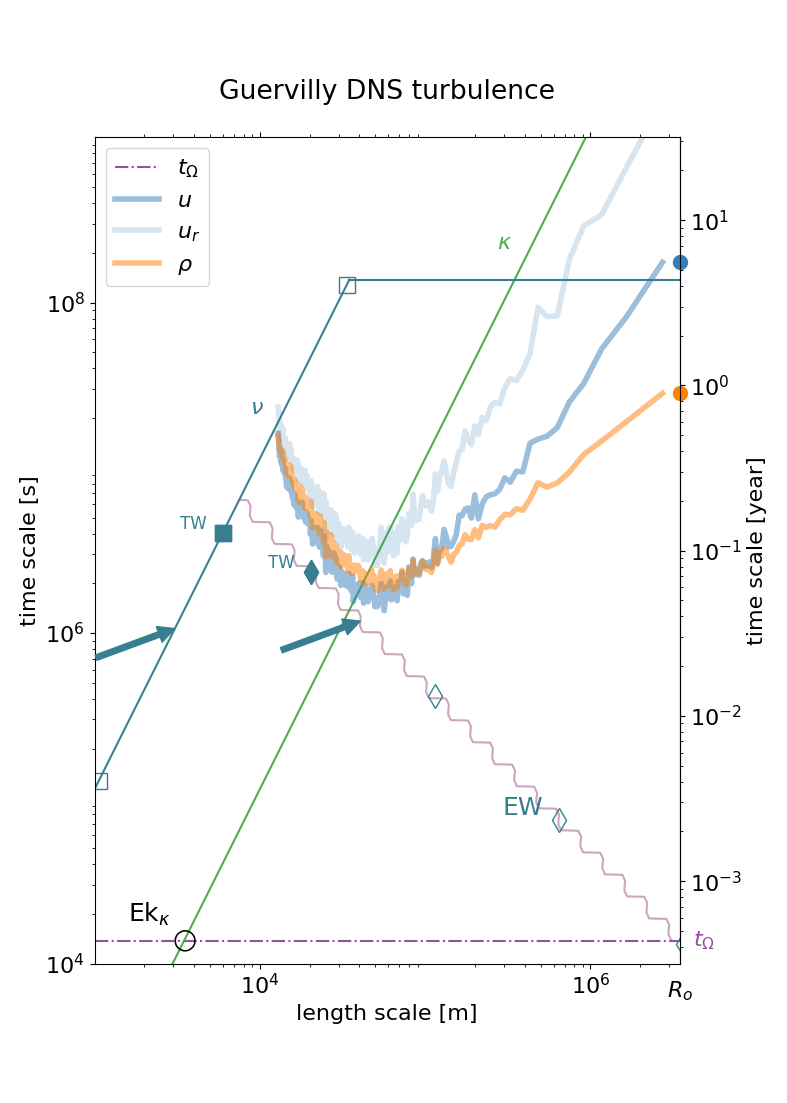}
			\caption{3D simulation from \citet{guervilly2019}.}
			\label{fig:E-8}
		\end{subfigure}
		\begin{subfigure}[b]{7.7cm}
			\includegraphics[width=\textwidth]{ 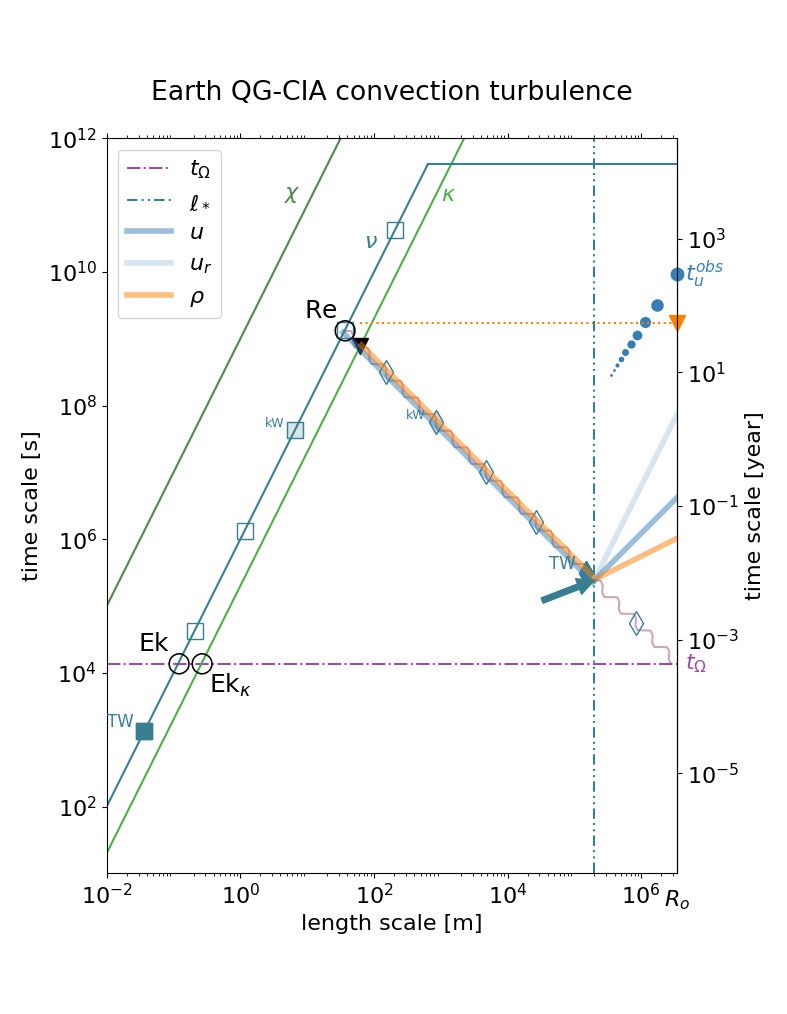}
			\caption{QG-CIA scenario for a non-magnetic Earth core.}
			\label{fig:QG-CIA}
		\end{subfigure}
		\caption{
		\tauell diagrams for non-magnetic rapidly rotating convection.
		Refer to Figure \ref{fig:convection_template} and section \ref{sec:phenomena} for a complete description of the background `template'.
		(a) \emph{3D numerical simulation from \citet{guervilly2019}.}
		Field variables $u$ and $\rho$ of the simulation are represented by blue $\tau_u(\ell)$ and orange $\tau_\rho(\ell)$ lines, respectively, which are \tauell translations of their respective volume-averaged order $m$-spectra (see Appendix \ref{sec:spectra_m}).
		Simulation's viscous dissipation in the bulk can be read on line $\tau_\nu(\ell)$ at the blue arrow, while  viscous dissipation in Ekman layers is marked by a blue arrow on the Rossby line.
		Additional pale blue line labeled $u_r$ gives the \tauell line of radial velocity, which gets much smaller than azimuthal velocity at large length-scales.
		(b) \emph{Scenario for the Earth assuming a QG-CIA force balance.}
		The available convective power $\mathcal{P}_{diss} \simeq 3$ TW sets time $\tau_*$ on the Rossby line (blue arrow) at which QG-CIA force balance yields the dominant vortex radius $\ell_*=\ell_\perp \simeq 200$ km, such that $\tau_u(\ell_\perp) = \tau_\rho(\ell_\perp) = \tau_{Rossby}(\ell_\perp)$.
		The QG-CIA balance governs flow at length-scales $\ell < \ell_*$ all the way to the intersections with diffusion lines $\tau_\kappa(\ell)$ and $\tau_\nu(\ell)$.
		At length-scales $\ell > \ell_*$, azimuthal flow velocities dominate over radial velocities (labeled $u_r$).
		}
	\end{figure}
\end{centering}

We focus here on rapid rotation regimes (small Rossby number), for which the leading order balance is quasi-geostrophic. Important results have been obtained for this regime by \citet{guervilly2019}, who performed numerical simulations of thermal convection at low Prandtl number ($\Pr = 10^{-2}$ and $10^{-1}$) in a sphere, at Ekman numbers down to $\Ek(R_o) = 10^{-8}$ in 3D, and to $\Ek(R_o) = 10^{-11}$ in quasi-geostrophic 2D.
In this section, we build the \tauell diagram of their most extreme 3D simulation, at $\Ek(R_o)=10^{-8}$and  $\Pr=10^{-2}$, and discuss its implications.
Figure \ref{fig:E-8} displays the \tauell diagram we obtain.

We first build its `template' as in Figure \ref{fig:convection_template}, using radius $R_o$ and $t_\Omega$ (spin rate's inverse) of the actual Earth's core as length-scale and timescale, respectively, in order to compare with the Earth.
Input dimensionless parameters $\Ek$ and $\Pr$ of the simulation provide values needed to build lines $\tau_\nu(\ell)=\ell^2/\nu$ and $\tau_\kappa(\ell)=\ell^2/\kappa$.
As in Figure \ref{fig:convection_template}, power dissipation markers are drawn along the Rossby and viscous lines, using outer core mass $M_o$.
The simulation provides the power dissipated by viscosity in the bulk, pointed by a blue arrow on the viscous line, and the slightly larger viscous dissipation in the Ekman boundary layer, pointed by a blue arrow on the Rossby line.

We now turn to extracting lines $\tau_u(\ell)$ and $\tau_\rho(\ell)$ from the simulation.
Line $\tau_u(\ell)$, labeled $u$, is obtained from the conversion of the volumetric average of a snapshot's azimuthal order $m$-kinetic energy spectrum, following equation (\ref{eq:tauell-m_u}) of Appendix \ref{sec:spectra_m}.
Note that line $\tau_u(\ell)$ stays above the Rossby line at all length-scales, implying that the flow is quasi-geostrophic, as expected from the high degree of $z$-invariance observed in this simulation \citep{guervilly2019}.
Remember that, in this case, length-scale $\ell$ has to be understood as the flow's length-scale in the equatorial plane.
Flow becomes anisotropic at large length-scale, as shown by the additional $\tau_{u_r}(\ell)$ line, labeled $u_r$, of radial velocities.
The azimuthal over radial velocity ratio increases with $\ell$.
Line $\tau_\rho(\ell)$, labeled $\rho$, is obtained in a similar way, using the conversion rule given by equation (\ref{eq:tauell-m_rho}) of Appendix \ref{sec:spectra_m}, with gravity given by $g = \frac{\Ra(R_o) R_o \, \kappa \nu}{R_o^4}$ and $\Ra(R_o)=2.5 \times 10^{10}$.

Both $\tau_u(\ell)$ and $\tau_\rho(\ell)$ lines display a sharp timescale minimum, very close to the Rossby line, defining length-scale $\ell_\perp$.
This simulation thus nicely illustrates the QG-CIA force balance, with  $\tau_{Rossby}(\ell_\perp) \sim \tau_u(\ell_\perp) \sim \tau_\rho(\ell_\perp)$, as advocated by \citet{guervilly2019}.
Note that the same force balance appears to apply for $\ell < \ell_\perp$, as envisioned by \citet{rhines1975}.
Length-scale $\ell_\perp$ coincides with the length-scale given by power dissipation occurring in Ekman layers (blue arrow pinned to the Rossby line).
As expected from equations (\ref{eq:tauQG}) and (\ref{eq:epsilon}), viscous dissipation is mostly due to flow at the minimum $\tau_u(\ell)$ time.

\subsection{\tauell diagram for a non-magnetic Earth core}
\label{sec:non-magnetic-Earth}

We now have all elements to start building a \tauell scenario for rotating convection in a non-magnetic Earth's core, which we present in Figure \ref{fig:QG-CIA}.
The goal is to create and draw realistic $\tau_u(\ell)$ and $\tau_\rho(\ell)$ lines over the background `template' of Figure \ref{fig:convection_template}.
We assume that viscous dissipation mainly occurs in Ekman layers.
Applying equation (\ref{eq:tauQG}), we obtain time $\tau_* = \left[ \frac{M_o R_o}{\mathcal{P}_{diss}} \sqrt{\frac{\nu}{\Omega^3}} \right]^{1/4}$, which yields the dissipated power $\mathcal{P}_{diss}$ that we estimate for the Earth's core (see section \ref{sec:core}), marked by a blue arrow in Figure \ref{fig:QG-CIA}.
We further formulate the ansatz of a QG-CIA force balance at the `optimum' length-scale $\ell_*$, which thus lies on the Rossby line.
We plot point ($\ell_*$, $\tau_*$) in Figure \ref{fig:QG-CIA}.

Inspired by Figure \ref{fig:E-8}, and by Rhines' arguments, we infer that flow obeys the QG-CIA balance for all length-scales $\ell < \ell_*$.
We thus plot $\tau_u(\ell)$ and $\tau_\rho(\ell)$ lines along the Rossby line all the way to their intersection with dissipation $\tau_\nu(\ell)$ and $\tau_\kappa(\ell)$ lines, respectively.
It remains to draw lines $\tau_u(\ell)$ and $\tau_\rho(\ell)$ for $\ell$ larger than $\ell_\perp$.
Flow becomes anisotropic for $\ell > \ell_\perp$, with radial velocities decreasing as $\ell$ increases, while azimuthal velocities increase with $\ell$.
For $\ell > \ell_\perp$ we thus loosely prescribe $\tau_{u_r}(\ell) \propto \ell^{2}$ (for a spectral energy density $E(k) \propto k$), $\tau_{u_{az}}(\ell) \simeq \tau_{u}(\ell) \propto \ell$ (for $E(k) \propto k^{-1}$), and $\tau_\rho(\ell) \propto \ell^{1/2}$ (for a $k^{-2}$-spectrum).

Reading the \tauell diagram of Figure \ref{fig:QG-CIA}, we see that core flow in a non-magnetic Earth would be quasi-geostrophic at all scales, with azimuthal velocities reaching 3 m s$^{-1}$, much larger than present-day core flow velocities represented by its $t_u^{obs}$ value and trend.
The radius of dominant columnar vortices would be around $200$ km.
Ekman layer viscous dissipation would dominate over bulk viscous dissipation by many orders of magnitude.

\section{\tauell regime diagrams for the Earth's core}
\label{sec:Earth}
In this section, we examine which \tauell regime diagrams to expect for the Earth's core.
Our goal is not to come up with an optimal or accurate scenario, but rather to illustrate how \tauell diagrams can help inventing and testing such scenarios.
We now consider the presence of a magnetic field and try to document Earth's core \tauell diagram, for which we presented a template in Figure \ref{fig:dynamo_template}.
Our starting point is the available convective power, as in section \ref{sec:non-magnetic-Earth}.
Building a turbulence scenario requires again estimating the balance of forces at different length-scales.

\subsection{\tauell expression of force balances in rotating convective dynamos}
\label{sec:force_balance_dynamo}
Following the approach of section \ref{sec:force_balance}, we add the Lorentz force in Navier-Stokes' equation:
\begin{equation}
	\rho \left( \partial_t \mathbf{u} +  \mathbf{u} \cdot \nabla \mathbf{u} + 2 \bm{\Omega} \times \mathbf{u} \right) = -\nabla p + \Delta \rho \mathbf{g}  + \mathbf{j} \times \mathbf{b} + \rho \nu \nabla^2 \mathbf{u},
	\label{eq:NS_dynamo}
\end{equation}
where $\mathbf{j}$ is the electric current density.
Following \citet{aurnou2017, aubert2019,schwaiger2019, schwaiger2020}, we consider two different situations: (i) Proudman-Taylor constraint is broken by the magnetic field and we get a three-term balance between Lorentz, buoyancy, and Coriolis (MAC force balance); (ii) Coriolis acceleration is dominant, and flow is quasi-geostrophic at leading order and obeys a QG-MAC force balance.

\subsubsection{Magneto-Archimedean-Coriolis (MAC)}
\label{sec:MAC}
When magnetic Lorentz force is strong enough to break quasi-geostrophy at scale $\ell_*$, one can get a balance between Lorentz, buoyancy and Coriolis forces, such that:
\begin{equation}
	\frac{\Omega u}{\ell_*} \sim \frac{b^2}{\rho \mu \ell_*^2} \sim \frac{\Delta \rho }{\rho} \frac{g}{\ell_*},
\end{equation}
analogous to the CIA balance with fluid velocity replaced by Alfv\'en wave velocity in the advection term.
Translating in \tauell language, we get:
\begin{equation}
	t_\Omega \tau_u(\ell_*) \sim \tau_b^2(\ell_*) \sim \tau_\rho^2(\ell_*),
	\label{eq:MAC_tauell}
\end{equation}
which implies:
\begin{equation}
	\Lambda_d(\ell_*) = \frac{t_\Omega \tau_u(\ell_*)}{\tau_b^2(\ell_*)} \sim 1.
\end{equation}
Graphically, this means that $\tau_b(\ell_*)$ and $\tau_\rho(\ell_*)$ both plot at mid-distance between $\tau_u(\ell_*)$ and $t_\Omega$.
This is the regime we get when the magnetic field is strong enough (see \citet{aurnou2017}).
The dynamical Elsasser number $\Lambda_d$ is of order one at scale $\ell_*$.

\subsubsection{Quasi-Geostrophic Magneto-Archimedean-Coriolis (QG-MAC)}
\label{sec:QG-MAC}
When the leading order force balance is quasi-geostrophic, the Coriolis term should only involve its ageostrophic part, at a length-scale $\ell_\parallel \sim R_o$.
QG-MAC balance at lenght-scale $\ell_\perp$ therefore writes:

\begin{equation}
	\frac{\Omega u}{\ell_\parallel} \sim \frac{b^2}{\rho \mu \ell_\perp^2} \sim \frac{\Delta \rho }{\rho} \frac{g}{\ell_\perp},
\end{equation}
analogous to the QG-CIA balance with fluid velocity replaced by Alfv\'en wave velocity in the advection term.
Translating in \tauell language, we get:
\begin{equation}
	\frac{R_o}{\ell_\perp}t_\Omega \tau_u(\ell_\perp) \sim \tau_\rho^2(\ell_\perp) \sim \tau_b^2(\ell_\perp),
	\label{eq:QG-MAC_tauell}
\end{equation}
which can also be written:
\begin{equation}
	\tau_{Rossby}(\ell_\perp) \, \tau_u(\ell_\perp) \sim \tau_\rho^2(\ell_\perp) \sim \tau_b^2(\ell_\perp).
	\label{eq:QG-MAC_tauell_alt}
\end{equation}
Graphically, this means that $\tau_b(\ell_\perp)$ and $\tau_\rho(\ell_\perp)$ both plot at mid-distance between $\tau_u(\ell_\perp)$ and $\tau_{Rossby}(\ell_\perp)$.

\subsection{\tauell diagram of a remarkable numerical simulation}
\label{sec:S2}
\begin{centering}
	\begin{figure}
		\begin{subfigure}[b]{6.8cm}
			\includegraphics[width=\linewidth]{ 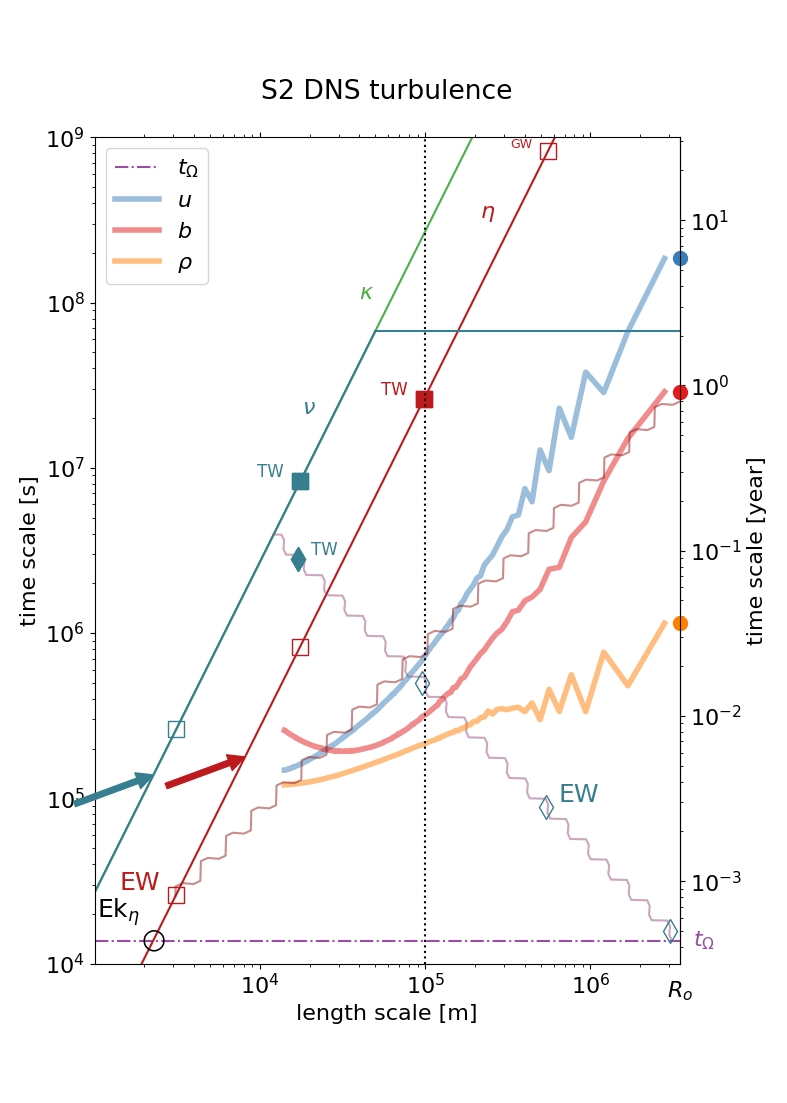}
			\caption{S2 DNS of \citet{schaeffer2017}.}
			\label{fig:S2}
		\end{subfigure}
		\begin{subfigure}[b]{7.7cm}
			\includegraphics[width=\linewidth]{ 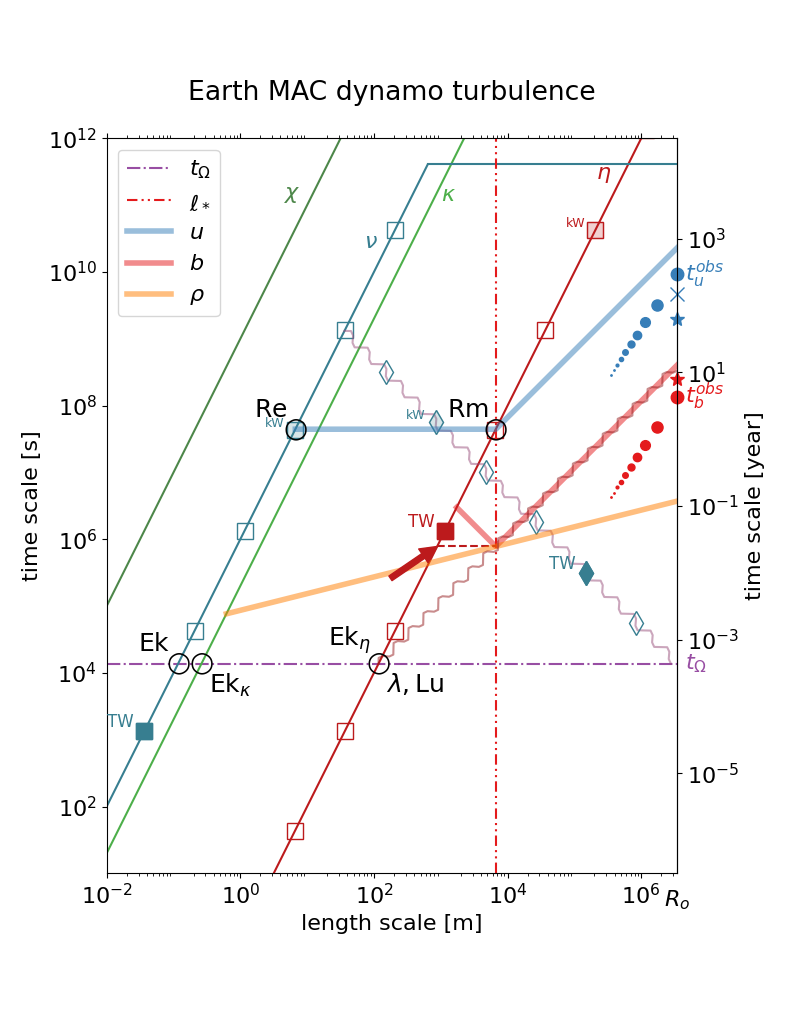}
			\caption{A simple MAC-balance scenario for Earth's core.}
			\label{fig:MAC}
		\end{subfigure}
		\caption{
		\tauell diagrams for rapidly rotating convective dynamo.
		Refer to Figure \ref{fig:dynamo_template} and section \ref{sec:phenomena} for a complete description of the background `template'.
		(a) \emph{3D numerical simulation S2 of \citet{schaeffer2017},}
		Field variables $u$, $b$ and $\rho$ of the simulation are represented by blue $\tau_u(\ell)$, red $\tau_b(\ell)$, and orange $\tau_\rho(\ell)$  lines, respectively, which are \tauell translations of their respective time- and volume-averaged degree $n$-spectra (see Appendix \ref{sec:spectra_n}).
		Simulation's viscous dissipation can be read on line $\tau_\nu(\ell)$ at the blue arrow, while Ohmic dissipation is marked by a red arrow on line $\tau_\eta(\ell)$.
		The black vertical dotted line marks the length-scale at which a QG-MAC force balance appears to be achieved.
		(b) \emph{Scenario for the Earth assuming a MAC force balance.}
		The available convective power $\mathcal{P}_{diss} \simeq 3$ TW sets time $\tau_*$ on the $\tau_\eta(\ell)$ line (red arrow).
		 Length-scale of maximum dissipation $\ell_* \simeq 7$ km is obtained by combining $\Rm(\ell_*) = 1$ and a MAC force balance at this same length-scale.
		Circles labeled $\Re$ and $\Rm$ at the intersections of line $\tau_u$ with lines $\tau_\nu(\ell)$ and $\tau_\eta(\ell)$ mark the scales at which the corresponding $\ell$-scale dimensionless number (see Table \ref{tab:dimensionless}) equals 1.
		Red and blue stars on the right $y$-axis mark magnetic intensity and velocity amplitude, respectively, predicted by \citet{christensen2006}'s scaling laws.
		Blue cross from \citet{davidson2013}'s velocity scaling law.
		}
		\label{fig:core}
	\end{figure}
\end{centering}

Let us start by building and discussing the \tauell regime diagram  (Figure \ref{fig:S2}) of one of the most extreme dynamo simulations available today: the S2 DNS of \citet{schaeffer2017}.

We first build its `template' as in Figure \ref{fig:dynamo_template}, using radius $R_o$ and $t_\Omega$ of the actual Earth's core as length-scale and timescale, respectively, in order to compare with the Earth.
Input dimensionless parameters of the simulation ($\Ek(R_o-R_i)=10^{-7}$, $\Pr=1$, $\Pm=0.1$) provide values needed to build lines $\tau_\nu(\ell)$, $\tau_\kappa(\ell)$ and $\tau_\eta(\ell)$.
As in Figure \ref{fig:dynamo_template}, power dissipation markers are drawn and labeled along the magnetic diffusion line, and along the Rossby and viscous lines, using outer core mass $M_o$.
Ohmic and viscous dissipations $D_\eta$ and $D_\nu$ of the simulation are obtained from Table 2 of \citet{schaeffer2017}, and scaled to Earth's core by: $\mathcal{P} = D \rho \nu^3/(R_o-R_i)$.
They are pointed by a red arrow on the magnetic diffusion line, and a blue arrow on the viscous line, respectively.

Next, we turn to extracting lines $\tau_u(\ell)$ and $\tau_b(\ell)$ from the simulation's spherical harmonic spectra, applying conversion rules of equations (\ref{eq:tauell_u}) and (\ref{eq:tauell_b}) of Appendix \ref{sec:spectra_n}, respectively.
The simulated acceleration of gravity $g$ at the top boundary is obtained from:
\begin{equation}
	g = \frac{\Ra^* R_o \, \kappa \nu}{(R_o-R_i)^4},
\end{equation}
with $\Ra^* = \Ra /\beta R_o = 2.4 \times 10^{13}$, where $\Ra$ is the classical large-scale Rayleigh number, and $\beta$ the imposed codensity gradient at the top boundary.
We then obtain line $\tau_\rho(\ell)$ by applying equation (\ref{eq:tauell_rho}) to the codensity spectrum multiplied by $\Pr^2$.

Reading the resulting \tauell diagram, we see that: magnetic energy largely dominates over kinetic energy ($\tau_b(R_o) \ll \tau_u(R_o)$); Ohmic dissipation dominates over viscous dissipation (compare dissipation powers indicated by arrows pinned to lines $\tau_\eta(\ell)$ and $\tau_\nu(\ell)$, respectively); flow should be largely quasi-geostrophic, since the $\tau_u(\ell)$ line stays above the Rossby line down to dissipation length-scales.
We also observe that both $\tau_u(\ell)$ and $\tau_b(\ell)$ lines have slopes steeper than 1 at large length-scale, while the slope of the $\tau_\rho(\ell)$ line is closer to 1/2.
Finally, we observe that a QG-MAC force balance seems approximately satisfied at length-scale $\ell_\perp \simeq R_o/18$, marked by a vertical dotted line, at which we observe $\frac{\tau_b}{t_\Omega} \frac{\tau_b}{\tau_u} \sim \frac{R_o}{\ell_\perp}
$ and $\tau_b(\ell_\perp) \sim \tau_\rho(\ell_\perp)$.

\subsection{A simple MAC-balance scenario}
\label{sec:MAC_scenario}
Figure \ref{fig:MAC} proposes a first attempt to complete the template of Figure \ref{fig:dynamo_template} with plausible $\tau_u(\ell)$,  $\tau_b(\ell)$ and  $\tau_\rho(\ell)$ lines.
As in section \ref{sec:non-magnetic-Earth}, our starting point is the available convective power $\mathcal{P}_{diss} \simeq 3$ TW.
We formulate the ansatz that it is dissipated by Joule heating only, which means that line $\tau_b(\ell)$ should get down to (but not below) time $\tau_* = \sqrt{M_o \eta / \mathcal{P}_{diss}}$ that yields a dissipation equal to $\mathcal{P}_{diss}$, as pointed by the red arrow on line $\tau_\eta(\ell)$.
However, in contrast with the situation of section  \ref{sec:non-magnetic-Earth}, we cannot attach line $\tau_b(\ell)$ to this point, because we anticipate that the velocity would be too weak at that scale to amplify the magnetic field.

One of the guidelines of our \tauell approach is that regime changes should occur where relevant $\ell$-scale dimensionless numbers reach 1, \textit{i.e.,} at the intersection of corresponding \tauell lines.
In a dynamo, we expect a regime change for $\Rm(\ell) \sim 1$, at the intersection of lines $\tau_u(\ell)$ and $\tau_\eta(\ell)$, below which magnetic diffusion takes precedence over induction.
Furthermore, if the magnetic field is strong, the intensity of turbulence is strongly reduced in that low $\Rm$ regime.
We thus assume that Ohmic dissipation is maximum at length-scale $\ell_*$ where $\Rm(\ell_*) = 1$, implying $\tau_b(\ell_*) \simeq \tau_*$.

This first condition links the velocity field to the magnetic field but is not sufficient to provide $\ell_*$.
Another constraint is needed, which we get from a force balance.
As a first guess, we request that our system obeys a MAC force balance at length-scale $\ell_*$.
Recalling equation \ref{eq:MAC_tauell}, we have:
\begin{equation*}
	t_\Omega \tau_u(\ell_*) \sim \tau_b^2(\ell_*) \sim \tau_\rho^2(\ell_*).
\end{equation*}
Together with condition $\Rm(\ell_*)=1$, this sets $\ell_*$ at a position such that $\tau_u(\ell_*) = \tau_\eta(\ell_*) = \tau_*^2/t_\Omega$, and implies $\tau_\rho(\ell_*) = \tau_*$, as drawn in Figure \ref{fig:MAC}.
Graphically, $\tau_*$ plots at mid-distance between $\tau_u(\ell_*)$ and $t_\Omega$.
We can also write:
\begin{equation}
	\left( \frac{\ell_*}{R_o} \right)^2 = \frac{\tau_*}{t_\Omega} \frac{\tau_*}{T_\eta} = \frac{1}{t_\Omega T_\eta^2} \frac{M_o R_o^2}{\mathcal{P}_{diss}},
	\label{eq:MAC_l_star}
\end{equation}
where $T_\eta \equiv \tau_\eta(R_o)$.

The next step consists in drawing lines $\tau_u(\ell)$ and $\tau_b(\ell)$ for $\ell < \ell_*$ and for $\ell > \ell_*$.
For $\ell < \ell_*$, we note that the regime of MHD turbulence in the presence of a strong magnetic field is characterized by steep energy density spectra: $E(k) \propto k^{-3}$ and $E_m(k) \propto k^{-5}$ \citep{alemany1979}.
Converting to \tauell language with equation (\ref{eq:tauell_new}), we get $\tau_u(\ell) \propto \ell^0$ and $\tau_b(\ell) \propto \ell^{-1}$, yielding the slopes drawn in Figure \ref{fig:MAC}.
Adding rotation further reduces the intensity of turbulence \citep{nataf2008, kaplan2018}, but we lack constraints on the resulting energy spectra.

In the dynamo region, for $\ell > \ell_*$, we assume $k^{-1}$ energy density spectra for both $u$ and $b$ (\textit{i.,e.}, $\tau(\ell) \propto \ell$), implying that line $\tau_b(\ell)$ follows the Alfv\'en wave line $\tau_{Alfven}(\ell)$.
This choice is mostly for pedagogical reasons as explained below.
Finally, we rather arbitrarily assume $\tau_\rho(\ell) \propto \ell^{1/2}$ at all length-scales.
This completes the \tauell diagram shown in Figure \ref{fig:MAC}.

Reading this \tauell diagram, we see that our scenario yields velocity and magnetic amplitudes that are not too far from the observed $t_u^{obs}$ and $t_b^{obs}$ values.
They translate into a magnetic to kinetic energy ratio of about $10^4$, according to equations (\ref{eq:E}) and (\ref{eq:E_m}). 
Bulk and boundary viscous dissipations have comparable amplitudes, both many orders of magnitude smaller than Ohmic dissipation, as assumed.
The smallest QG vortices (on the Rossby line) are very sluggish, with a turnover time of about 1 year and a radius of 1 km.
Magnetic diffusion is largest at a length-scale of about 10 km.

It is interesting to observe that our scenario implies that the Alfv\'en line intersects line $\tau_\eta(\ell)$ at time $t_\Omega$, meaning that all three $\ell$-scale dimensionless numbers $\Ek_\eta$, $\Lu$ and $\lambda$ equal 1 at this same scale.
This implies that $\tau_b(R_o)$ can be deduced from the intersection of $\tau_\eta(\ell)$ and $t_\Omega$ lines.
In other words, even though our scenario has been built to achieve a given Ohmic dissipation $\mathcal{P}_{diss}$, the actual value of $\tau_b(R_o)$ does not depend on $\mathcal{P}_{diss}$.
Only the kinetic energy depends on $\mathcal{P}_{diss}$, as given by these scaling laws:
\begin{eqnarray}
	\tau_u(R_o) &=& \left( \frac{T_\eta \tau_*^2 }{t_\Omega} \right)^{1/2}
	= \left( \frac{\Omega M_o R_o^2}{\mathcal{P}_{diss}} \right)^{1/2} 
	\label{eq:MACu}\\
	\tau_b(R_o) &=& \left( t_\Omega T_\eta \right)^{1/2}
	= \frac{R_o}{\sqrt{\Omega \eta}}.
	\label{eq:MACb}
\end{eqnarray}

\subsection{A simple QG-MAC balance scenario}
\label{sec:QG-MAC_scenario}
\begin{centering}
	\begin{figure}
		\begin{subfigure}[b]{=7.7cm}
			\includegraphics[width=\textwidth]{ 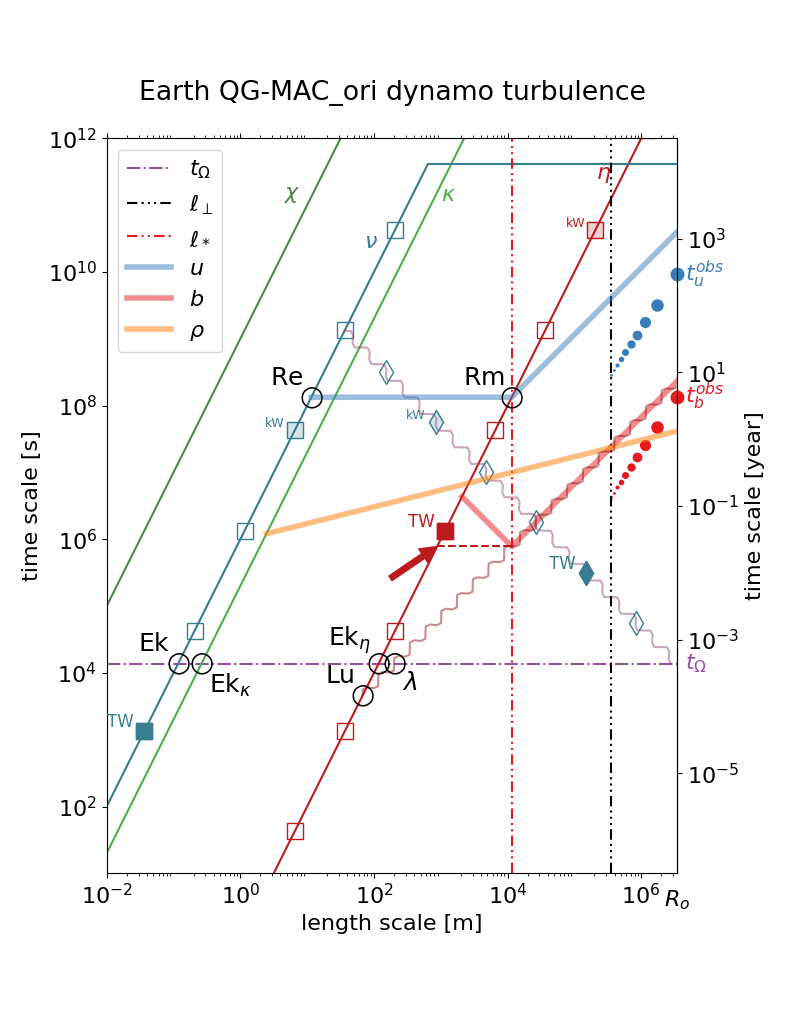}
			\caption{Simple QG-MAC balance scenario.}
			\label{fig:QG-MAC}
		\end{subfigure}
		\begin{subfigure}[b]{7.7cm}
			\includegraphics[width=\linewidth]{ 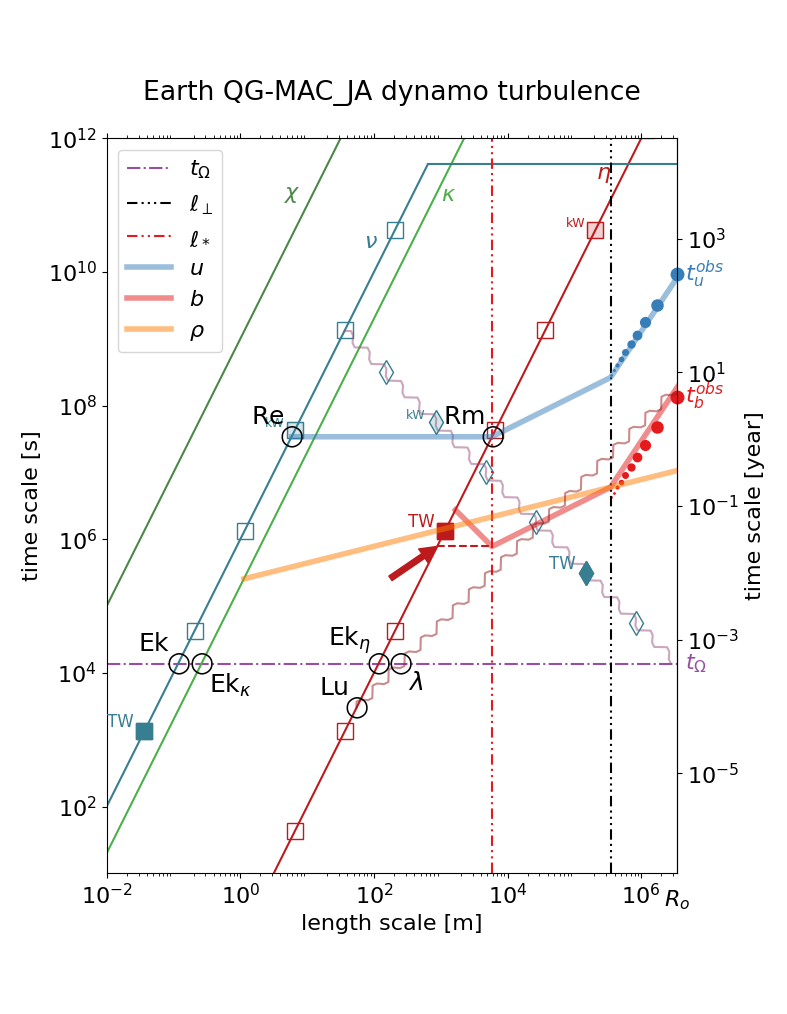}
			\caption{QG-MAC balance scenario \textit{\`a la Aubert}}
			\label{fig:QG-MAC_JA}
		\end{subfigure}
		\caption{
		\tauell diagrams of two QG-MAC force balance scenarios for the Earth core.
		Refer to Figure \ref{fig:dynamo_template} and section \ref{sec:phenomena} for a complete description of the background `template', and to Figure \ref{fig:MAC} for more details.
		Both scenarios assume that the available convective power $\mathcal{P}_{diss} \simeq 3$ TW is dissipated by Joule effect, setting time $\tau_*$ marked by a red arrow on the magnetic diffusion line, and that the flow obeys a QG-MAC balance at a prescribed length-scale $\ell_\perp=R_o/10$ indicated by a blue vertical dot-dash line.
		(a) \emph{a simple QG-MAC balance scenario.} It is assumed that $\tau_u(\ell)$ and $\tau_b(\ell)$ are proportional to $\ell$ between $\ell_*$ and $R_o$ (see section \ref{sec:QG-MAC_scenario}).
		(b) \emph{a QG-MAC balance scenario \`a la Aubert.} This scenario assumes $\tau_u(\ell) \propto \ell^{1/2}$ for $\ell_* < \ell < \ell_\perp$ (invariant vorticity) and $\tau_u(\ell) \propto \ell^{3/2}$ for $\ell_\perp < \ell < R_o$ (following the trend of observations $t_u^{obs}$). Same trends for $\tau_b(\ell)$ (see section \ref{sec:QG-MAC_JA_scenario}).
		}
		\label{fig:QG-MACs}
	\end{figure}
\end{centering}

The simple MAC-balance scenario of the previous section meets a problem: line $\tau_u(\ell)$ plots above the Rossby line for all length-scales $\ell > \ell_*$, and all these scales have $\Lambda_d(\ell) < 1$.
This means that flow should be quasi-geostrophic in this scale range, even though the magnetic field is strong.
Leading-order force balance should therefore be quasi-geostrophic, with convective motions forming columnar vortices parallel to the spin axis.
Figure \ref{fig:QG-MAC} shows a first plausible alternative scenario, built as follows.

As in section \ref{sec:MAC_scenario} we first obtain the minimum magnetic time $\tau_b$, which is set to $\tau_* = \sqrt{M_o \eta / \mathcal{P}_{diss}}$, marked by a red arrow on line $\tau_\eta(\ell)$, and the corresponding length-scale of maximum dissipation $\ell_*$ is such that $\Rm(\ell_*)=1$.
This first condition links the velocity field to the magnetic field but is not sufficient to provide $\ell_*$.
Another constraint is needed, which we get from a QG-MAC force balance.
We recall that such a balance at length-scale $\ell_\perp$ is given by equation (\ref{eq:QG-MAC_tauell}): 
\begin{equation*}
	\frac{R_o}{\ell_\perp}t_\Omega \tau_u(\ell_\perp) \sim \tau_b^2(\ell_\perp) \sim \tau_\rho^2(\ell_\perp).
\end{equation*}

Graphically, this means that $\tau_b(\ell_\perp)$ plots at mid-distance between $\tau_u(\ell_\perp)$ and $\tau_{Rossby}(\ell_\perp)$.
One then needs to guess at which scale $\ell_\perp$ this balance should apply.
In contrast with our simple MAC scenario, one cannot choose $\ell_\perp = \ell_*$, because this would place $\tau_u(\ell_*)$ very far down, below the $t_\Omega$ line, breaking quasi-geostrophy, and yielding $\mathcal{E}_k \gg \mathcal{E}_m$ in strong disagreement with observations.
And we also need to decide how velocity and magnetic fields vary between $\ell_*$ and $R_o$.
As in section \ref{sec:MAC_scenario} we assume $\tau_u(\ell) \propto \ell$ and $\tau_b(\ell) \propto \ell$.
We then obtain:
\begin{equation}
	\left( \frac{\ell_*}{R_o} \right)^3 = \left(  \frac{\ell_\perp}{R_o} \right)^2 \frac{\tau_*}{t_\Omega} \frac{\tau_*}{T_\eta} .
	\label{eq:QG-MAC_l_star}
\end{equation}

Figure \ref{fig:QG-MAC} displays the \tauell diagram of such a QG-MAC scenario with $\ell_\perp = R_o/10$.
Comparing with Figure \ref{fig:MAC}, we see that this scenario predicts a larger magnetic over kinetic energy ratio, with $\tau_u$ above the Rossby line down to scales of a few hundred meters.
Another difference is the level of line $\tau_\rho(\ell)$.

\subsection{A QG-MAC balance scenario \`a la Aubert}
\label{sec:QG-MAC_JA_scenario}
In the previous scenario, choosing $\ell_\perp = R_o/10$ was borrowed from \citet{aubert2017} and \citet{aubert2019}, who find it in good agreement with their numerical simulation results.
Following \citet{davidson2013}, \citet{aubert2019} proposes a $\tau_u(\ell)$ scaling for  $\ell_* < \ell < \ell_\perp $ that differs from the one we used in section \ref{sec:QG-MAC_scenario}.
In that interval, \citet{davidson2013} infers that vorticity is independent of $\ell$.
This translates into $\tau_u(\ell) \propto \ell^{1/2}$ instead of $\tau_u(\ell) \propto \ell$.
Using the same scaling for $\tau_b(\ell)$, length-scale $\ell_*$ is then given by:
\begin{equation}
	\left( \frac{\ell_*}{R_o} \right)^{5/2} = \left(  \frac{\ell_\perp}{R_o} \right)^{3/2} \frac{\tau_*}{t_\Omega} \frac{\tau_*}{T_\eta} .
	\label{eq:JA_l_star}
\end{equation}

The corresponding \tauell regime diagram is shown in Figure \ref{fig:QG-MAC_JA}, where we chose to follow the observed $\ell^{3/2}$ trend for $\tau_u$ and $\tau_b$ above $\ell_\perp$ (blue and red disks, respectively).
This scenario provides an amazing fit to the observed $t_u^{obs}$ and $t_b^{obs}$ values.
Furthermore, we observe that the dynamical Elsasser number $\Lambda_d(\ell) = \frac{t_\Omega}{\tau_b(\ell)}\frac{\tau_u(\ell)}{\tau_b(\ell)}$ remains below 1 at all length-scales $\ell$, validating our assumption of leading-order Quasi-Geostrophy.

This scenario yields the following (ugly-looking) scaling laws for the largest scale velocity and magnetic fields:
\begin{eqnarray}
	\tau_u(R_o) &=& \left( \frac{R_o}{\ell_\perp} \right)^{1/10} \left( \frac{T_\eta^2 \tau_*^6 }{t_\Omega^3} \right)^{1/5}
	\label{eq:QG-MAC_JAu}\\
	\tau_b(R_o) &=& \left( \frac{R_o}{\ell_\perp} \right)^{13/10} \left( t_\Omega T_\eta \tau_*^3 \right)^{1/5}.
	\label{eq:QG-MAC_JAb}
\end{eqnarray}

\subsection{The interesting case of Venus}
\label{sec:Venus}
The internal structure of Venus is very poorly known, but we know that it does not generate a detectable magnetic field.
This important difference from its sister planet Earth is classically explained by a different thermal history, leading to a hot mantle convecting beneath a rigid lid, preventing core cooling, hence halting the convective engine of the dynamo \citep{stevenson1983,nimmo2002}.

\begin{centering}
	\begin{figure}
		\includegraphics[width=7.7cm]{ 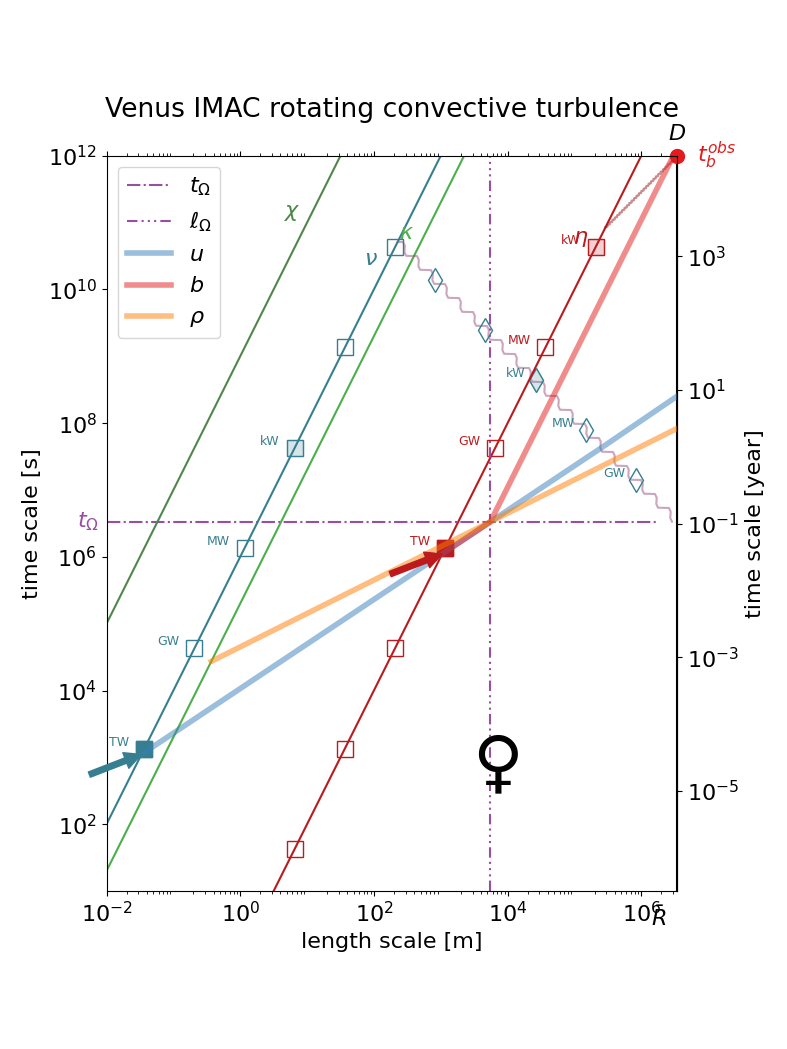}
		\caption{
		Devil's advocate \tauell regime diagram for the core of Venus.
		All relevant properties are assumed identical to that of the Earth's core (see section \ref{sec:core}), except for the spin rate (rotation period of 243 days).
		Venus `template' is built as in Figure \ref{fig:dynamo_template}.
		Value $t_u^{obs} \simeq 10^{12}$ s is deduced from an upper bound on Venus' undetected magnetic field. 
		We speculate that the difference in rotation period has a dramatic impact on the dynamo, and section \ref{sec:Venus} proposes a tentative IMAC force balance scenario, which results in the displayed \tauell diagram. 
		}
		\label{fig:Venus}
	\end{figure}
\end{centering}

Venus and Earth also differ by their spinning rate: one turn in 243 days instead of one day.
This difference is usually considered as unimportant since rotation still appears overwhelming, with an Ekman number $\Ek(R_o) \sim 10^{-13}$ \citep{russell1980}.
However, if we adopt for Venus the same physical properties as for Earth, including its available convective power $\mathcal{P}_{diss}$, but update the spinning rate to the one of Venus, we meet a problem, illustrated in Figure \ref{fig:Venus}. 

We first observe that Ohmic dissipation time $\tau_*= \sqrt{M_o \eta / \mathcal{P}_{diss}}$ plots below the $t_\Omega$ line.
We also see that the core of Venus would not be able to dissipate such a power by friction in its Ekman layers, as indicated by the markers along the Rossby line (compare with Figure \ref{fig:QG-CIA}).
Venus does not appear to be a `fast rotator', and we should not expect the flow to be dominantly Quasi-Geostrophic as in Earth's core.
It might be more similar to the solar dynamo.

We therefore propose a very tentative `devil's advocate' scenario, in which we assume that dissipation takes place in the bulk, with equipartioned viscous and Ohmic dissipations, yielding the blue arrow on the viscous line, and the red arrow on the magnetic diffusion line.
In this region far below the $t_\Omega$ line, Kolmogorov energy density spectra $E(k) \propto k^{-5/3}$ seem plausible for both kinetic and magnetic energies, which translate into $\tau_u(\ell) \propto \ell^{2/3}$ and $\tau_b (\ell) \propto \ell^{2/3}$, as drawn in Figure \ref{fig:Venus} from their respective dissipation scales up to timescale $t_\Omega$, which is met at a length-scale $\ell_\Omega \simeq 5$ km.

The force balance we expect there is of type Inertia-Magneto-Archimedean-Coriolis (IMAC), given by:
\begin{equation}
	t_\Omega \tau_u(\ell_\Omega) \sim \tau_u^2(\ell_\Omega) \sim \tau_b^2(\ell_\Omega)\sim \tau_\rho^2(\ell_\Omega),
	\label{eq:IMAC_tauell}
\end{equation}
which we use to plot $\tau_\rho(\ell_\Omega)$.
At length-scales larger than $\ell_\Omega$ the IMAC scenario drawn in Figure \ref{fig:Venus} assumes $\tau_\rho(\ell) \propto \ell^{1/2}$, $\tau_u(\ell) \propto \ell^{2/3}$, and $\tau_b(\ell) \propto \ell^2$.
The unusual very large $\ell$ exponent for the magnetic field, corresponding to an $E(k) \propto k$ magnetic energy density spectrum, predicts an undetectable magnetic field at the largest length-scale.
Such a spectrum is indeed observed at the surface of the Sun \citep{finley2023}, where the magnetic field is much stronger at small length-scales than at large length-scales. 

Our exercise is very formal since we recall that there are good reasons to believe that such a $\mathcal{P}_{diss}$ convective power is not available for Venus \citep{stevenson1983,nimmo2002}.
Nevertheless, our diversion to Venus questions the notion of `fast rotator' and shows that for a given convective power $\mathcal{P}_{diss}$, planet's spin rate might play a more important role than simply inferred from the low Ekman number it delivers.

\section{Discussion}
\label{sec:discussion}
In this section, we try to show how \tauell diagrams can help building a new vision on several ongoing debates, such as the validity domain of various scaling laws and the controversy on the dominant length-scale in the core.
We also illustrate the link with `path strategies' and propose extensions of this approach.

\subsection{Validity domain of dynamo regimes}
One goal of our approach is to better appreciate the validity domain of the various regimes encountered in planetary cores.
For example, no dynamo will exist if line $\tau_u(\ell)$ plots too high in the diagram, yielding $\Rm(R_o) \sim 1$.
In Earth's core, this would only happen for $\mathcal{P}_{diss}$ values several orders of magnitude smaller than our reference value of $3$ TW.
Actually, very low values (even negative) are not excluded by some thermal history models, before the birth of the inner core (\textit{e.g.,} \citet{landeau2022}).

At the other end, large $\mathcal{P}_{diss}$ values can pull the dynamo generation domain ($\Rm(\ell) > 1$) partly beneath the Rossby line.
Complete columnar vortices won't then have time to form at length-scales below the Rossby line, and we might have a somewhat different turbulence regime.
\citet{davidson2014} proposed an original dynamo scenario that corresponds to such a situation.
Noting that inertial waves are strongly helical, and that flow helicity is a key ingredient for dynamo action, he suggests that Earth's dynamo might operate this way.
None of the three scenarios we presented (Figures \ref{fig:MAC}, \ref{fig:QG-MAC}, \ref{fig:QG-MAC_JA}) puts $\Rm(\ell) \sim 1$ below the Rossby line.
However, it would only take $\mathcal{P}_{diss} \simeq 10$ TW for the QG-MAC scenario \textit{\`a la Aubert} (Figure \ref{fig:QG-MAC_JA}) to qualify (remember we only consider orders of magnitude).

An even more dramatic regime change should occur when $\tau_* < t_\Omega$, as in our devil's advocate scenario for Venus (Fig. \ref{fig:Venus}).
Then, the object should not be considered as a `fast rotator' anymore, and the dynamo regime is more of a small-scale type.
This occurs when $\mathcal{P}_{diss} > M_o \eta \Omega^2$.
For the Earth's core, this translates into a power of $10^4$ TW, which was never reached in Earth's history.


We believe that \tauell diagrams could help inferring relevant dynamo regimes for planets and stars.

\subsection{Dynamo scaling laws}
\label{sec:scaling_laws}
\citet{christensen2010b} nicely reviews a number of scaling laws proposed to infer the magnetic field intensity of planets (and stars).
Among the nine proposed scaling laws he lists, five relate magnetic intensity to planetary rotation rate $\Omega$, with no influence of the available convective power $\mathcal{P}_{diss}$, three involve both  $\Omega$ and $\mathcal{P}_{diss}$, and one only $\mathcal{P}_{diss}$.
The latter one, first proposed by \citet{christensen2006}, stems from the analysis of a large corpus of numerical simulations of the geodynamo, backed by an appraisal of the dominant terms in the governing equations.
It is the law preferred by \citet{christensen2010b} who shows that it is in good agreement with the measured magnetic intensity of planets and stars.

\noindent
Translated in \tauell formulation, Christensen's preferred laws yield:
\begin{eqnarray}
	\tau_u(R_o) &=&\frac{1}{c_u} \left( \frac{M_o R_o^2 \Omega^{1/2}}{\tilde{F} \, \mathcal{P}_{diss}} \right)^{2/5}
	\simeq \left( \frac{T_\eta^2 \tau_*^4}{t_\Omega}\right)^{1/5}
	\label{eq:CAu}\\
	\tau_b(R_o) &=& \frac{1}{\sqrt{2 c_b f_{ohm}}} \left( \frac{M_o R_o^2}{\tilde{F} \,  \mathcal{P}_{diss}} \right)^{1/3}
	\simeq \left( T_\eta \tau_*^2 \right)^{1/3},
	\label{eq:CAb}
\end{eqnarray}
from equations (32) and (15) of \citet{christensen2010b}.
Both large-scale flow velocity $U \equiv R_o/\tau_u(R_o)$ and magnetic field $B  \equiv (\rho \mu)^{-1/2} R_o/\tau_b(R_o)$ do not depend upon magnetic diffusivity\footnote{Note that one can define a dissipation time $T_{diss} = \sqrt[3]{\frac{M_o R_o^2}{\mathcal{P}_{diss}}}$, which makes it more explicit that $T_\eta \tau_*^2= T_{diss}^3$ is independent of magnetic diffusivity $\eta$.}, and $B$ is also independent of spin rate $\Omega$.
These predictions are displayed by blue and red stars on the right axis in Figure \ref{fig:MAC}.
While $B$-prediction agrees well with observations, $U$-prediction over-estimates flow velocity.

\citet{davidson2013} re-examined this question and noted that inertia is playing too strong a role in \citet{christensen2006}'s simulations.
He derived a QG-MAC-balance variant and obtained a revised flow velocity law (his equation (15)), which translates into:
\begin{equation}
	\tau_u(R_o) \simeq \Omega^{1/3}  \left( \frac{M_o R_o^2 }{\mathcal{P}_{diss}} \right)^{4/9}
	= \left( \frac{T_\eta^4 \tau_*^8}{t_\Omega^3}\right)^{1/9},
	\label{eq:D13u}\\
\end{equation}
in which $U$ is independent of magnetic diffusivity $\eta$.
Davidson's prediction is drawn as a blue $\times$ in Figure \ref{fig:MAC}.

Our discussion on scenario validity domain questions the obtention of universal dynamo scaling laws.
However, we have seen that such scaling laws can be easily derived from the \tauell diagram of various scenarios.
Each scenario produces $\tau_u(R_o)$ and $\tau_b(R_o)$ that combine $\tau_*$ (or $T_{diss}$), $T_\eta$ and $t_\Omega$ with various powers.
Equations (\ref{eq:MACu}) and (\ref{eq:MACb}) give the scaling laws for a simple MAC scenario.
Although built to dissipate a given convective power $\mathcal{P}_{diss}$ by Ohmic dissipation, this scenario provides a large-scale magnetic field $B$ that does not depend upon $\mathcal{P}_{diss}$, while $U$ is independent of magnetic diffusivity $\eta$.
The prediction of our QG-MAC scenario `\`a la Aubert' is given by equations (\ref{eq:QG-MAC_JAu}) and (\ref{eq:QG-MAC_JAb}) for $U$ and $B$, respectively.


\subsection{Dominant length-scale controversy}
\label{sec:length-scale}
Several recent studies have revived a debate on the dominant length-scales to be expected in the Earth's core \citep{yan2021, bouillaut2021,yan2022, cattaneo2022, nicoski2024, abbate2023, hawkins2023}.
Fast rotation inhibits thermal convection, and viscosity is needed to break the Taylor-Proudman constraint \citep{chandrasekhar1961}.
Studies on the onset of thermal convection in rapidly rotating plane layers \citep{chandrasekhar1961}, annulus \citep{busse1970}, and spherical shells or spheres \citep{busse1970, jones2000, dormy2004} therefore all show that, at the onset of convection, the flow consists in columnar vortices aligned with the rotation axis, with a width $\ell_c \sim R_o [\Ek(R_o)]^{1/3}$ (see Appendix \ref{sec:onset}).
In the Earth's core, such columns would be ~30 m in diameter, extending several thousand kilometers in length.
At least two effects could strongly widen such columns: (i) the magnetic field \citep{chandrasekhar1961, fearn1979, sreenivasan2011, aujogue2015}; (ii) non-linear flow advection \citep{rhines1975, ingersoll1982, gillet2007b, jones2015, guervilly2019}.

The effect of the magnetic field can be studied at the onset of convection, but only for given -often simplistic- geometries of the magnetic field that never occur in natural systems.
These studies show that convective cells can get as large as the depth of the container $R_o$ when a magnetic field is present.
Concerning planetary dynamos, one can argue that such an effect can only be invoked when the dynamo has already produced a strong enough magnetic field.
This led \citet{cattaneo2022} to call on the Moon for help, arguing that in an unmagnetized Earth, ``convection would consist of hundreds of thousands of extremely thin columns'' that would be unable to produce a magnetic field.

This view is clearly challenged by \citet{guervilly2019} who advocate a dominant convective length-scale around 30 km.
Their prediction rests on the extrapolation of a QG-CIA force balance to Earth's core conditions, and is backed by numerical simulations such as the one we present in Figure \ref{fig:E-8}.
Our QG-CIA scenario, shown in Figure \ref{fig:QG-CIA} obeys the same rules.
It predicts an even larger dominant length-scale ($\sim 200$ km) because it is built on the available convective power $\mathcal{P}_{diss}$ rather than on the observed large-scale velocity, which is reduced by the magnetic field.
Interestingly, Figure \ref{fig:E-8} shows that the simulations of \citet{guervilly2019} remain very close to the convection onset, which would yield a critical length-scale $\ell_c$ close to the intersection of the $\tau_\kappa(\ell)$ and $\tau_{Rossby}(\ell)$ lines (see Appendix \ref{sec:onset}).
Yet, they reach a turbulent QG-CIA balance because of the low value of the Prandtl number ($\Pr=10^{-2}$) that allows for large Reynolds numbers near the onset \citep{aubert2001, kaplan2017}.

Our \tauell diagrams clearly emphasize the role of $\mathcal{P}_{diss}$ in controlling the dominant length-scale, both with and without a magnetic field.
This is particularly clear in Figure \ref{fig:QG-CIA} where the minima of curves $\tau_u(\ell)$ and $\tau_\rho(\ell)$ would `slide' along the Rossby line following the $\mathcal{P}_{diss}$ blue arrow.
This diagram also suggests that while viscosity in the bulk controls the convective length-scale at the onset (black triangle), it progressively looses its importance as viscous dissipation in the Ekman layers takes over.
To achieve a turbulent convection regime where viscosity in the bulk is unimportant, it is crucial to provide another way of dissipating the convective power, like Ekman layer friction or Ohmic dissipation, which are both present in Earth's core.
Because you always need to dissipate the power injected into turbulence, the previous sentence sounds rather obvious.
Yet, this might explain why several recent studies find no evidence of a diffusion-free scaling, in which viscosity plays no role.
In the numerical simulations of \citet{yan2021, yan2022, nicoski2024}, boundaries are stress-free so that no other energy sink than viscous dissipation in the bulk can equilibrate the convective power.

\subsection{Path strategy}
\label{sec:path}
Our \tauell approach has much in common with the `path strategy' devised by Julien Aubert and colleagues \citep{aubert2017, aubert2021, aubert2022, aubert2023}.
Their strategy also targets the dynamical regime of a given object, the Earth's core, and aims at defining a `path' that numerical simulations should follow to reach that regime.
As stressed by \citet{dormy2016}, minimizing $\Pm$ at a given Ekman number $\Ek(R_o) = \nu / \Omega R_o^2$ is not a good strategy, since it increases magnetic Ekman number $\Ek_\eta(R_o) = \eta / \Omega R_o^2$, while \citet{christensen2010a} show that $\Ek_\eta(R_o) < 10^{-4}$ is needed for obtaining Earth-like geodynamo simulations.
Hence the need for targeting a more elaborate `distinguished limit' \citep{dormy2016}.

Aubert and his colleagues stress the importance of respecting the hierarchy of dynamic times, and propose a unidimensional path based on scaling laws obtained by \citet{davidson2013} for a QG-MAC force balance.
Position along the path is measured by an $\epsilon$ parameter.
As exposed in \citet{aubert2023}, the idea is to keep the magnetic diffusion time $T_\eta = \tau_\eta(R_o)$ fixed to the Earth value, and to  decrease the magnetic Ekman number step by step, by decreasing the rotation time $t_\Omega$ from values at which appropriate numerical simulations are available (defining $\epsilon=1$) down to the actual value of the Earth ($\epsilon=10^{-7}$).
Long-period behaviour of the Earth's core is well recovered from $\epsilon=1$, while decreasing $\epsilon$ to Earth's value allows for the development of short-term dynamics.

\begin{centering}
	\begin{figure}
		\begin{subfigure}[b]{=5cm}
			\includegraphics[width=\textwidth]{ 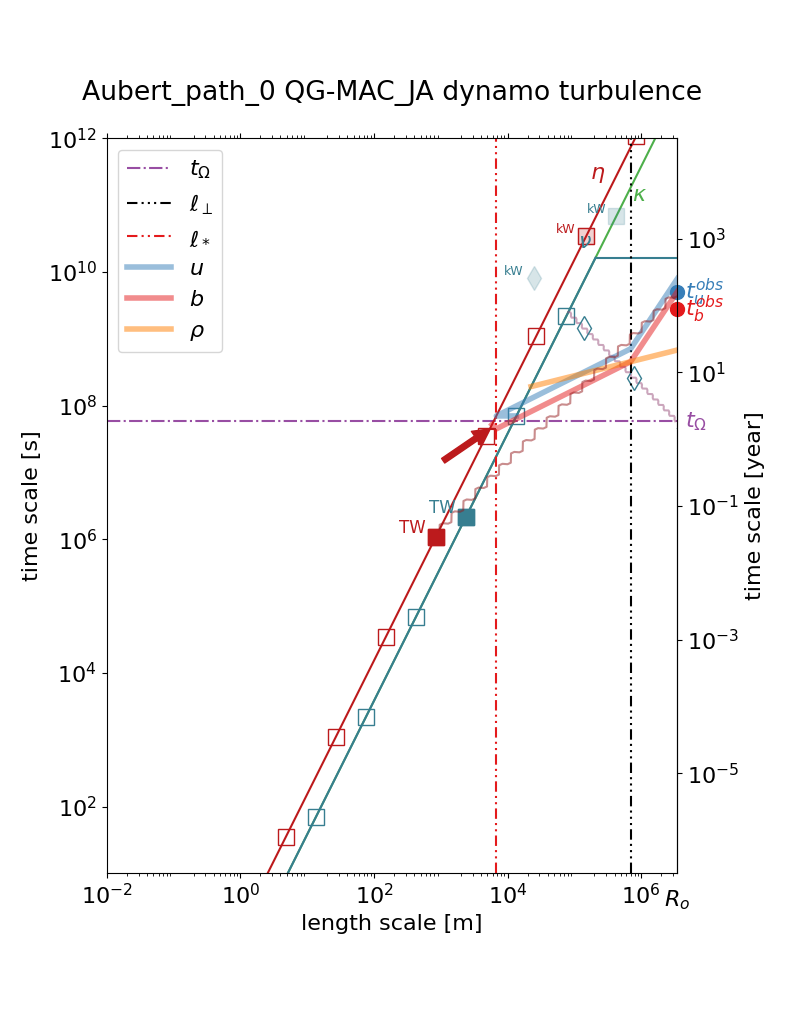}
			\caption{Path 0\%.}
			\label{fig:path_0}
		\end{subfigure}
		\begin{subfigure}[b]{5cm}
			\includegraphics[width=\linewidth]{ 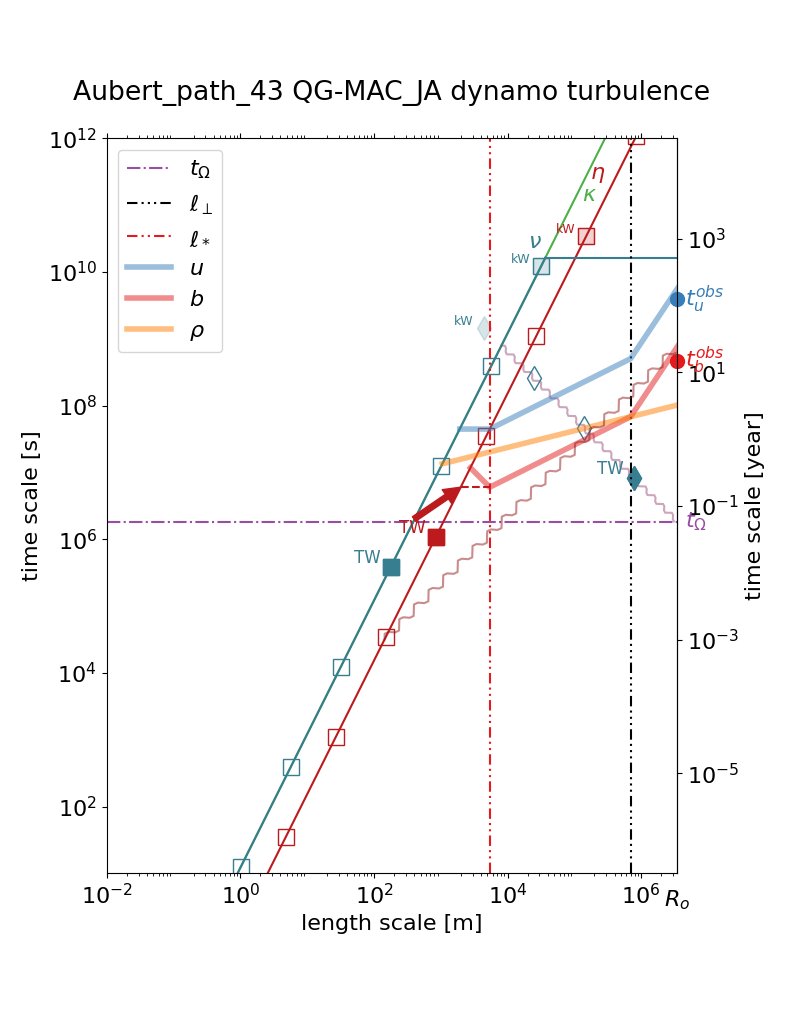}
			\caption{Path 43\%.}
			\label{fig:path_43}
		\end{subfigure}
		\begin{subfigure}[b]{5cm}
			\includegraphics[width=\linewidth]{ 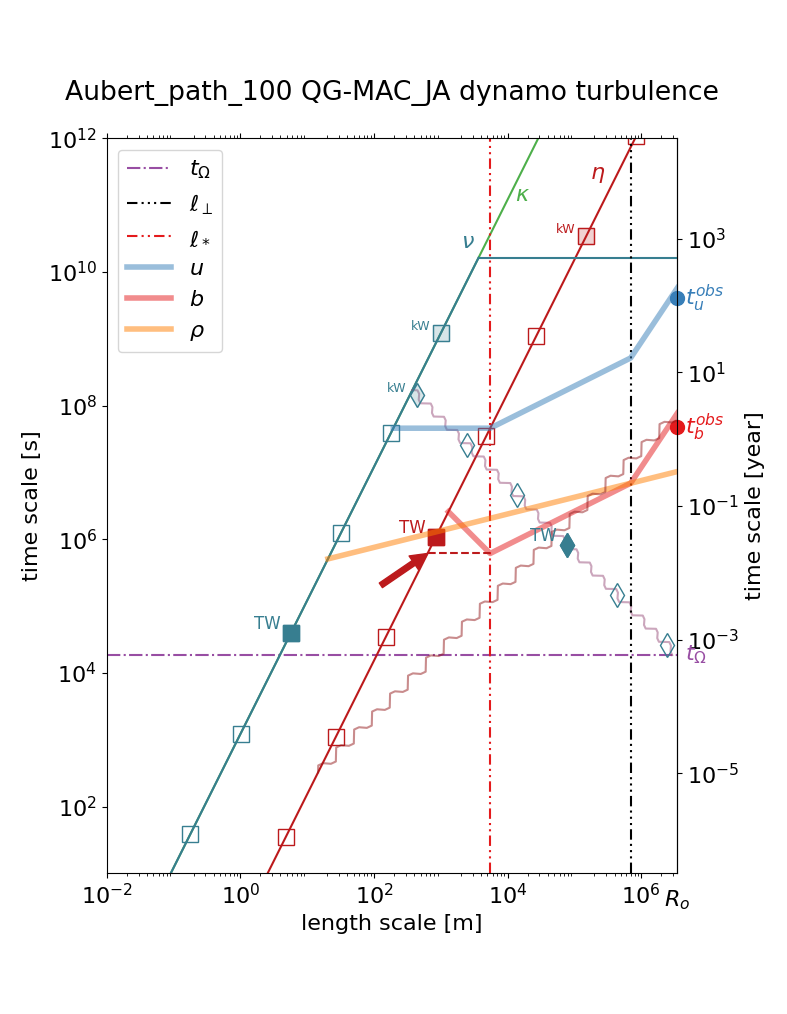}
			\caption{Path 100\%.}
			\label{fig:path_100}
		\end{subfigure}
		\caption{\tauell regime diagrams illustrating the `path strategy' devised by Aubert and colleagues. (a) 0\% of path ($\epsilon = 1$). (b) 43\% of path  ($\epsilon = 10^{-3}$).  (b) 100\% of path  ($\epsilon = 10^{-7}$).
		See text for explanations.
		}
		\label{fig:path}
	\end{figure}
\end{centering}

Figure \ref{fig:path} shows how \tauell diagrams can help visualize and extend this strategy.
We show three diagrams, which correspond to three different values of $\epsilon$: $1$, $10^{-3}$, and $10^{-7}$.
Their `template' are built from the values of the magnetic Ekman and Prandtl numbers listed for `prior PB' in Table 1 of \citet{aubert2023}.
The values of $\mathcal{P}_{diss}$ (red arrow), $t_u^{obs}$ (blue disk), and $t_b^{obs}$ (red disk), are from the corresponding simulation outputs given in his Table 2.
We then add the $\tau_\rho(\ell)$, $\tau_u(\ell)$ and $\tau_b(\ell)$ lines that our `\`a la Aubert' QG-MAC balance scenario (see section \ref{sec:QG-MAC_JA_scenario}) predicts for the listed $\mathcal{P}_{diss}$.
An amazing fit of the `observed' $t_u$ and $t_b$ is obtained, choosing $\ell_\perp = R_o/5$.

A key contribution of Aubert and colleagues was to determine how the input parameters of numerical simulations should scale with $\epsilon$ in order to follow the path.
This is where the QG-MAC scaling laws of \citet{davidson2013} come in, and this is also where \tauell diagrams can help.
Using the scaling laws obtained for $\tau_u(R_o)$ and $\tau_b(R_o)$ in section \ref{sec:QG-MAC_JA_scenario}, we test two different paths.
Note that parameter $\epsilon$ is defined such that $\Ek_\eta[\epsilon] = \sqrt{\epsilon} \, \Ek_\eta[1]$, implying  $t_\Omega[\epsilon] = \sqrt{\epsilon} \, t_\Omega[1]$.

Aubert and colleagues chose to keep $\Rm(R_o)$ constant along the path, \textit{i.e.,} $\tau_u(R_o)[\epsilon] = C^{te}$.
Equation (\ref{eq:QG-MAC_JAu}) then implies $\tau_*[\epsilon] \propto \sqrt{t_\Omega[\epsilon]} \propto \epsilon^{1/4}$, since $T_\eta$ does not vary with $\epsilon$, and assuming that $\ell_\perp/R_o$ is constant as well.
We thus get $\mathcal{P}_{diss}[\epsilon] \propto \epsilon^{-1/2}$ since $\tau_* = \sqrt{M_o \eta / \mathcal{P}_{diss}}$.
Equation (\ref{eq:QG-MAC_JAb}) then implies $\tau_b(R_o)[\epsilon] \propto \epsilon^{1/2}$.
These scalings agree with those of \citet{aubert2023}, as can be seen in Figure \ref{fig:path}.

Instead of keeping $\Rm(R_o)$ constant along the path, one could choose to keep the dynamo-generating domain above the Rossby line, as we obtain for the Earth.
Imposing that $\Rm(\ell)=1$ occurs at the intersection of lines $\tau_\eta(\ell)$ and $\tau_{Rossby}(\ell)$ yields $\tau_u(R_o)[\epsilon] \propto \epsilon^{1/4}$, $\tau_b(R_o)[\epsilon] \propto \epsilon^{3/8}$, and $\mathcal{P}_{diss}[\epsilon] \propto \epsilon^{-11/12}$.
It would be interesting to run numerical simulations along such a path.

\section{Limitations and perspectives}
\label{sec:limitations}
Let us first recall that our \tauell approach is \emph{not} a theory of turbulence.
We try to formulate plausible scenarios by identifying scales at which a change in turbulence regime should occur, and by patching scaling laws appropriate for each regime.
We thus entirely depend on the availability of such laws, which can be brought by experiments, theory, and numerical simulations.

Our approach contains a fair number of assumptions and approximations.
How realistic are the conversion rules we employ to `translate' force balances and turbulent spectra in \tauell language?
Does the minimum dynamic time control dissipation?
What controls length-scale $\ell_\perp$ that we had to guess for writing QG-MAC-type force balances?

For simplicity reasons, we have treated planetary cores as simple full spheres.
The application to actual planets requires to at least consider spherical shells of various thicknesses instead.
An extension to giant planets and stars also requires taking into account compressibility and free-slip boundaries.

Our results suggest that dissipation of Quasi-Geostrophic flows in Ekman layers takes over bulk dissipation in non-magnetic rapidly rotating convection when it gets strongly super-critical.
Can it be tested? How smooth is the transition?
Our devil's advocate scenario for Venus suggests a transition to a very different regime if dissipation is too large to be taken up by Ekman friction.
Is there evidence for such a transition?
How sharp is it?


We observe that density anomaly spectra from numerical simulations are rarely displayed, while they convey valuable information.
We also note that spectra from laboratory experiments are scarce (but see \citet{madonia2023}) and too often given in `arbitrary units', preventing their conversion into \tauell representation.
We are lacking experimental data on turbulence for rotating convection in a sphere in presence of strong magnetic fields.

\tauell diagrams provide hints on how velocity and magnetic field scale with length-scale.
This might be useful for observers who need such constraints to tune their magnetic field and core-flow inversions \citep{gillet2015, baerenzung2016}.
As a matter of fact, our analysis suggests that the flat spherical harmonic spectra observed at low degrees $n$ for both flow velocity and magnetic field cannot extend much beyond degree 10 without meeting dissipation problems.

\section{Conclusion}
\label{sec:conclusion}
\tauell regime diagrams are a simple graphical tool that proves useful for inventing or testing dynamic scenarios for planetary cores.
Tradition in fluid dynamics is to characterize systems by dimensionless numbers, usually based on `typical' large-scale quantities.
Past decades have seen large efforts to develop a more detailed description of phenomena that operate at different scales.
This has led to the apparition of even more dimensionless numbers, in which the various scales involved do not always figure very clearly, and to the construction of somewhat unintelligible scaling laws.
By defining $\tau$ timescales that depend on $\ell$ length-scales over their entire range, we hope to make these choices more explicit.
By providing a simple graphical identity to these scales, we wish to make their meaning more intuitive.
Contrary to spectra in `arbitrary units', \tauell diagrams give insight into regimes and balances which are paramount to rotating, magnetized and/or stratified fluids, where waves can be present and significantly alter the dynamics.

Because they put together most key properties of a given object, \tauell regime diagrams constitute a nice identity card.
We think this applies to numerical simulations and laboratory experiments as well.
Both approaches enable extensive parameter surveys, which are crucial for exploring and understanding different regimes.
Being object-oriented, \tauell diagrams are not easily applied to such surveys, but we think they would very valuably complement classical scaling law plots.
The idea would be to draw \tauell diagrams for a few representative members and end-members of the survey, which would nicely illustrate their validity range.

Our article thus has two goals.
The first goal is to provide all ingredients for building your own \tauell diagram, be it of a numerical simulation, a laboratory experiment or theory.
To that end, we included construction rules, examples, technical appendices, and Python scripts (supplementary material).
The second goal is to demonstrate the potential of \tauell regime diagrams for suggesting and testing various scenarios for Earth's dynamo.

Convinced that available convective power $\mathcal{P}_{diss}$ is a key control parameter, and the one that can most readily be estimated for other planets and exoplanets, we have modified our original approach \citep{nataf2015} to propose and discuss a few scenarios built upon this input data.
This results in a more challenging exercise, calling for force balance inspection.
We show that the \tauell translation of relevant force balances is very handy and telling.

We built several geodynamo scenarios to test MAC (Figure  \ref{fig:MAC}) and QG-MAC (\ref{fig:QG-MAC}, \ref{fig:QG-MAC_JA}) force balances.
The validity domain of these scenarios shows up well in \tauell diagrams.
A QG-MAC scenario `\`a la Aubert' looks particularly appealing, and could have applied to the Earth over its entire history.
We note that in such a scenario, flow in the dynamo-generating region remains Quasi-Geostrophic, with a dynamical Elsasser smaller than 1, even though the magnetic to kinetic energy ratio is of order $10^4$.
In contrast, Venus would have a hard time entering that regime, because of its slow rotation (Figure \ref{fig:Venus}).
This calls for a re-analysis of what is called a `fast rotator'.

\tauell regime diagrams also help us addressing several on-going debates, such as the the validity of various scaling laws, and the question of the dominant convective length-scale in the Earth's core.
We speculate that dissipation in Ekman layers drives non-magnetic rapidly rotating convection towards a QG-CIA force balance when the flow is turbulent enough, promoting dominant length-scales much larger than the iconic $R_o \, \Ek^{1/3}$ length-scale at convection onset (Figure \ref{fig:QG-CIA}).

We use \tauell diagrams to illustrate the concept of `path strategy' developed by Julien Aubert and colleagues (Figure \ref{fig:path}), and we propose an interesting alternative to their original path.

\section*{Conflicts of interest}
The authors declare no competing financial interest.

\section*{Dedication}
The manuscript was mostly written by HCN. NS ran all presented numerical simulations. Both
authors have given approval to the final version of the manuscript.

\section*{Acknowledgments}
We thank the French Academy of Science and Electricit\'e de France for granting their ``Amp\`ere Prize'' to our ``Geodynamo'' team, and B\'ereng\`ere Dubrulle for early encouragements. HCN thanks Peter Davidson, Julien Aubert, Thomas Gastine, Franck Plunian, Sacha Brun, Antoine Strugarek and Quentin Noraz for useful discussions, and Emmanuel Dormy for his review. The authors would like to thank the Isaac Newton Institute for Mathematical Sciences, Cambridge, for support and hospitality during the programme ``Frontiers in dynamo theory: from the Earth to the stars'' where work on this paper was undertaken. This work was supported by EPSRC grant no EP/R014604/1.
ISTerre is part of Labex OSUG@2020 (ANR10 LABX56).
Our manuscript greatly benefited from the thorough reviews of three anonymous referees (including rather harsh but stimulating comments by a `lengths and time scales person').

\CDRGrant[EPSRC]{EP/R014604/1}

\section*{Supplementary data}
Supplementary material is available in document Nataf\_Schaeffer\_SupMat.pdf.
All Python scripts and data used to produce the figures of this article are available in package tau-ell\_programs.zip at
article's URL 
and at \href{https://gricad-gitlab.univ-grenoble-alpes.fr/natafh/shell_tau-ell_programs}{https://gricad-gitlab.univ-grenoble-alpes.fr/natafh/shell\_tau-ell\_programs}.

\appendix
\section{\tauell representation of turbulent spectra}
\label{sec:spectra}
Our `fuzzy physics' approach targets gathering dominant physical phenomena in a common frame.
Some of them are classically defined in wavenumber or frequency space, rather than in physical space, hence the need for some conversion.
This appendix lists the different types of energy spectra that are usually obtained from observations, numerical simulations and experiments, and derives recipes for converting them into \tauell language.

%

\subsection{Choosing a conversion rule}
\label{sec:conversion}
We consider different expressions of total energy per unit mass $\mathcal{U}$:
\begin{equation}
	\mathcal{U} = \int{E(k) dk} \hspace{0.3cm} \text{ or }  \hspace{0.3cm} \mathcal{U} = \sum_{k_i}{E}_B(k_i)  \hspace{0.3cm} \text{ or }  \hspace{0.3cm} \mathcal{U}(r) = \frac{1}{\rho}\sum_{n}\mathcal{L}_b(n,r).
\end{equation}
$E(k)$ is the classical \emph{spectral energy density} introduced by \citet{kolmogorov1941}.
$E_B(k_i)$ is the discrete equivalent of $E(k)$ for flow `in a box' \citep{lesieur2008}.
$\mathcal{L}_b(n,r)$ is the spherical harmonic degree $n$ component at radius $r$ of Lowes-Mauersberger spectrum widely used in geomagnetism \citep{lowes1966}.
Note that we denote $n$ the spherical harmonic degree, often denoted $l$ or $\ell$, in order to avoid confusion with our length-scale $\ell$.

A flow with spectral energy density $E(k) \propto k^{-5/3}$ yields the same $k$-exponent for its discrete energy spectrum $E_B$ \citep{lesieur2008,stepanov2014}, and a $n^{-5/3}$ Lowes-Mauersberger spectrum.
However, pre-factors may differ.
More importantly, the conversion of energy spectra into \tauell equivalents is questionable.
Indeed, no exact conversion between spectral energy density and eddy velocity can be drawn, as thoroughly discussed by \citet{davidson2005}.

In \citet{kolmogorov1941}'s universal turbulence, an \emph{eddy turnover time} is classically derived as:
\begin{equation}
	\tau_u(\ell) \simeq \ell^{3/2} \left[E(k) \right] ^{-1/2}  \hspace{0.3cm}  \text{with} \hspace{0.3cm} \ell \simeq 1/k,
	\label{eq:tauell_K41}
\end{equation}
where $\ell$ is the `size' of the eddy.
A similar result is obtained using \emph{velocity increments} $S_2(\ell) = <[u(x+\ell)-u(x)]^2> = C_2 (\epsilon \ell)^{2/3}$.
This approach is appealing for Kolmogorov-type self-similar inertial range where no length-unit other than $\ell$ should appear, and where large eddies are more energetic than small eddies.
This conversion rule was used in \citet{nataf2015}.

In our object-oriented approach, the integral length scale $R_o$ plays a role, and when translating numerical simulation energy spectra in \tauell form, we were thus tempted to simply define $\tau_u(\ell)$ as $\ell(n) \left[\mathcal{S}_u(n) \right] ^{-1/2}$ where $\mathcal{S}_u(n)$ is the degree $n$ component of the spherical harmonic spectrum of $u^2$, and $\ell(n)$ is given by:
\begin{equation}
	\ell(n) \simeq \frac{1}{2} \frac{\pi R_o}{n+1/2}.
	\label{eq:ell_of_n}
\end{equation}
It turns out that this choice is not consistent with our \tauell representation, in which the $n^{-5}$ spectrum of a system in Rhines' equilibrium at all scales should translate into $\tau_u(\ell) \propto \ell^{-1}$ and plot along the Rossby line.

We thus use conversion rules similar to that of equation (\ref{eq:tauell_K41}), such as:
\begin{equation}
	\tau_u(\ell(n)) \simeq \ell(n) \left[n \, \mathcal{S}_u(n) \right] ^{-1/2}  \hspace{0.3cm}  \text{with} \hspace{0.3cm} \ell(n) \simeq \frac{1}{2} \frac{\pi R_o}{n+1/2}.
	\label{eq:tauell_new}
\end{equation}

%
%

\subsection{Application to various relevant spectra}
\label{sec:application}
We now detail the \tauell conversion of spectra commonly obtained from observations, numerical experiments and laboratory experiments.

\subsubsection{Lowes-Mauersberger spectrum}
\label{sec:spectra_LM}
In geomagnetism, the variation of magnetic energy with length-scale is usually measured by its Lowes-Mauersberger spectrum \citep{lowes1966}.
This spectrum is expressed in terms of Gauss coefficients $g_n^m$ and $h_n^m$ of scalar magnetic potential $V$, which defines the internal magnetic field at any radius above the core-mantle boundary when the mantle is considered as an electrical insulator.

Potential $V(r,\theta,\varphi)$ is then solution of Laplace equation and can be expressed in terms of spherical harmonics as:
\begin{equation}
	V(r,\theta,\varphi) = R_o \sum_{n=1}^\infty \sum_{m=0}^n \left( \frac{R_o}{r}\right)^{n+1} (g_n^m \cos m \varphi + h_n^m \sin m \varphi) P_n^m(\cos \theta),
\end{equation}
where $P_n^m$ are the Schmidt semi-normalized associated Legendre functions of degree $n$ and order $m$.

Following \citet{langlais2014}, the Lowes-Mauersberger spectrum at any $r>R_o$ is then given by the suite of $\mathcal{L}_b(n)$ defined by:
\begin{equation}
	\mathcal{L}_b(n,r) = (n+1) \sum_{m=0}^n ((g_n^m)^2 + (h_n^m)^2) \left( \frac{R_o}{r}\right)^{2n+4}.
\end{equation}
The total magnetic energy per unit mass at radius $r$ is obtained as:
\begin{equation}
	\mathcal{U}_m(r) \equiv \frac{B^2(r)}{2 \rho \mu} =  \frac{1}{\rho}\sum_{n=1}^\infty \mathcal{L}_b(n,r).
	\label{eq:Btot}
\end{equation}
Spherical harmonic degree $n$ is related to our $\ell$ length-scale by:
 \begin{equation}
	\ell(n) \simeq \frac{1}{2} \frac{\pi R_o}{n+1/2}.
	\label{eq:n2ell}
\end{equation}
We deduce from equations (\ref{eq:Btot}) and (\ref{eq:n2ell}) that the magnetic field $b(\ell,r)$ at length-scale $\ell$ and radius $r$ is given by:
 \begin{equation}
 	b(\ell(n),r) \simeq \sqrt{2 \mu \mathcal{L}_b(n,r)}.
\end{equation}
We finally obtain $\tau_b(\ell)$ from:
 \begin{equation}
 	\tau_b(\ell(n)) \simeq \frac{\ell(n)}{\sqrt{2 n \mathcal{L}_b(n) / \rho}} \hspace{0.3cm} \text{with}  \hspace{0.3cm} \ell(n) = \frac{1}{2} \frac{\pi R_o}{n+1/2}.
	\label{eq:tauell_LM}
\end{equation}
A flat Lowes-Mauersberger spectrum (constant $\mathcal{L}_b(n)$), such as observed for the Earth's magnetic field at the core-mantle boundary, thus translates into $\tau_b(\ell) \propto \ell^{3/2}$.

\subsubsection{Degree $n$-spectra from numerical simulations}
\label{sec:spectra_n}
Numerical simulations of planetary dynamos are most often performed using a pseudo-spectral  expansion in spherical harmonics $Y_n^m(\theta, \varphi)$ of degree $n$ and order $m$.
Degree-$n$ or order-$m$ spectra are thus readily obtained for both velocity, magnetic and codensity fields.
These spectra are usually for $u^2$, $b^2$ and $C^2$ in dimensionless units, such that the sum over all $n$ and $m$ yields $2/\rho$ times the energy per unit mass of that dimensionless field.
Given length-scale $\mathrm{L}$ and time-scale $\mathrm{T}$ chosen in the simulation, $u$ and $b$ spectra should be multiplied by $\mathrm{L}^2/\mathrm{T}^2$ (assuming that $b$ is expressed in Alfv\'en wave velocity units).
 
Given these precisions, the procedure is similar to that exposed in Appendix \ref{sec:spectra_LM}.
$\ell(n)$ is still given by:
 \begin{equation}
 	\ell(n) = \frac{1}{2} \frac{\pi R_o}{n+1/2}.
	\label{eq:tauell_ell}
\end{equation}
One should keep in mind that for a given degree $n$, the corresponding length-scale varies linearly with radius.
This is ignored in our approach.
If $\mathcal{S}_x(n)$ is the $n$-element of the dimensionless $x^2$ spectrum, then the corresponding $\tau_x(\ell)$ lines for $x = u$, $b$ and $\rho$ are given by:
\begin{eqnarray}
 	\tau_u(\ell(n)) & \simeq & \frac{\mathrm{T}}{\mathrm{L}} \frac{\ell(n)}{\sqrt{n \mathcal{S}_u(n)}},
	\label{eq:tauell_u} \\
 	\tau_b(\ell(n)) & \simeq & \frac{\mathrm{T}}{\mathrm{L}} \frac{\ell(n)}{\sqrt{n \mathcal{S}_b(n)}},
	\label{eq:tauell_b} \\
 	\tau_\rho(\ell(n)) & \simeq & \sqrt{\frac{\ell(n)}{g \, \sqrt{n \mathcal{S}_\rho(n)}}},
	\label{eq:tauell_rho}
\end{eqnarray}
where gravity $g$ (in dimensional units) is obtained from the input Rayleigh number.

In the example of Figure \ref{fig:E-8} from \citet{guervilly2019}, the simulated acceleration of gravity $g$ at the top boundary is obtained from:
\begin{equation}
	g = \frac{\Ra R_o \, \kappa \nu}{R_o^4},
\end{equation}
with $\Ra = 2.5 \times 10^{10}$.

In the example of Figure \ref{fig:S2} from \citet{schaeffer2017}, $g$ is obtained from:
\begin{equation}
	g = \frac{\Ra^* R_o \, \kappa \nu}{(R_o-R_i)^4},
\end{equation}
with $\Ra^* = \Ra/\beta R_o = 2.4 \times 10^{13}$, where $\beta$ is the imposed codensity gradient at the top boundary.
We then obtain the $\tau_\rho(\ell)$ line by applying equation (\ref{eq:tauell_rho}) to the codensity spectrum multiplied by $\Pr^2$.

\subsubsection{Order $m$-spectra from numerical simulations}
\label{sec:spectra_m}

Quasi-geostrophic vortices are better characterized by their order-$m$ spectra $\mathcal{S}_x(m)$ than by their degree-$n$ spectra, in particular in 2D QG simulations.
Thus \citet{guervilly2019} display $m$-spectra of their 3D and QG convection simulation results.
Translation into \tauell is obtained as in Appendix \ref{sec:spectra_n}, replacing $\mathcal{S}_x(n)$ by $\mathcal{S}_x(m)$:
\begin{eqnarray}
 	\tau_u(\ell(m)) & \simeq & \frac{\mathrm{T}}{\mathrm{L}} \frac{\ell(m)}{\sqrt{m \mathcal{S}_u(m)}},
	\label{eq:tauell-m_u} \\
 	\tau_b(\ell(m)) & \simeq & \frac{\mathrm{T}}{\mathrm{L}} \frac{\ell(m)}{\sqrt{m \mathcal{S}_b(m)}},
	\label{eq:tauell-m_b} \\
 	\tau_\rho(\ell(m)) & \simeq & \sqrt{\frac{\ell(m)}{g \, \sqrt{m \mathcal{S}_\rho(m)}}},
	\label{eq:tauell-m_rho}
\end{eqnarray}
with:
 \begin{equation}
 	\ell(m) = \frac{1}{2} \frac{\pi R_o}{m+1/2}.
	\label{eq:tauell-m_ell}
\end{equation}

For the simulation presented in Figure \ref{fig:E-8}, $\mathcal{S}_x(m)$ and $\mathcal{S}_x(n)$ spectra are very similar, apart for even-odd oscillations in the $n$-spectra due to equatorial symmetry.

\subsubsection{Frequency spectrum}
\label{sec:spectra_exp}
In laboratory experiments, turbulent spectra are more easily obtained from signal $x(\mathbf{r},t)$ measured in the time-domain ($t=0$ to $T$) at a given position $\mathbf{r}$.
Power spectral density (PSD) is then computed from its Fourier transform $\hat{x}_T(\mathbf{r},f)$ as:
\begin{equation}
	\tilde{E}(\mathbf{r},f) = \lim_{T \to \infty} \frac{1}{T} | \hat{x}_T(\mathbf{r},f) |^2.
	\label{eq:PSD}
\end{equation}

When a mean flow $\mathbf{U(\mathbf{r})}$ is present, and when turbulence is weak enough, a time record of velocity reflects the advection of the spatial variation of velocity \citep[e.g.,][]{frisch1995}.
Extensions to intense turbulence have been developed \citep{pinton1994}.
Taylor's hypothesis \citep{taylor1938} then permits to obtain a kinetic energy density wavenumber spectrum $E(\mathbf{r},k)$ from the velocity frequency power spectrum $\tilde{E}(\mathbf{r},f)$:
\begin{equation}
	E(\mathbf{r}, k(f)) = \frac{U}{2 \pi}\tilde{E}(\mathbf{r}, f) \hspace{0.3cm}  \text{with}  \hspace{0.3cm} k(f) = \frac{2 \pi f}{U},
	\label{eq:PSDtok}
\end{equation}
where $U = ||\mathbf{U(\mathbf{r})}||$.
Strictly speaking, this is a $k_{\sslash}$-spectrum, valid for wavenumber $\mathbf{k}$ parallel to mean velocity vector $\mathbf{U}(\mathbf{r})$.
It is the same as a $k$-spectrum for isotropic turbulence.
The \tauell translation is then obtained from equation (\ref{eq:tauell_K41}).

Note that Taylor's hypothesis cannot be applied to magnetic spectra unless magnetic diffusion is small enough for the frozen flux approximation to apply.

%
%

\section{\tauell diagram for convection onset}
\label{sec:onset}
\begin{centering}
	\begin{figure}
		\includegraphics[width=7.7cm]{ 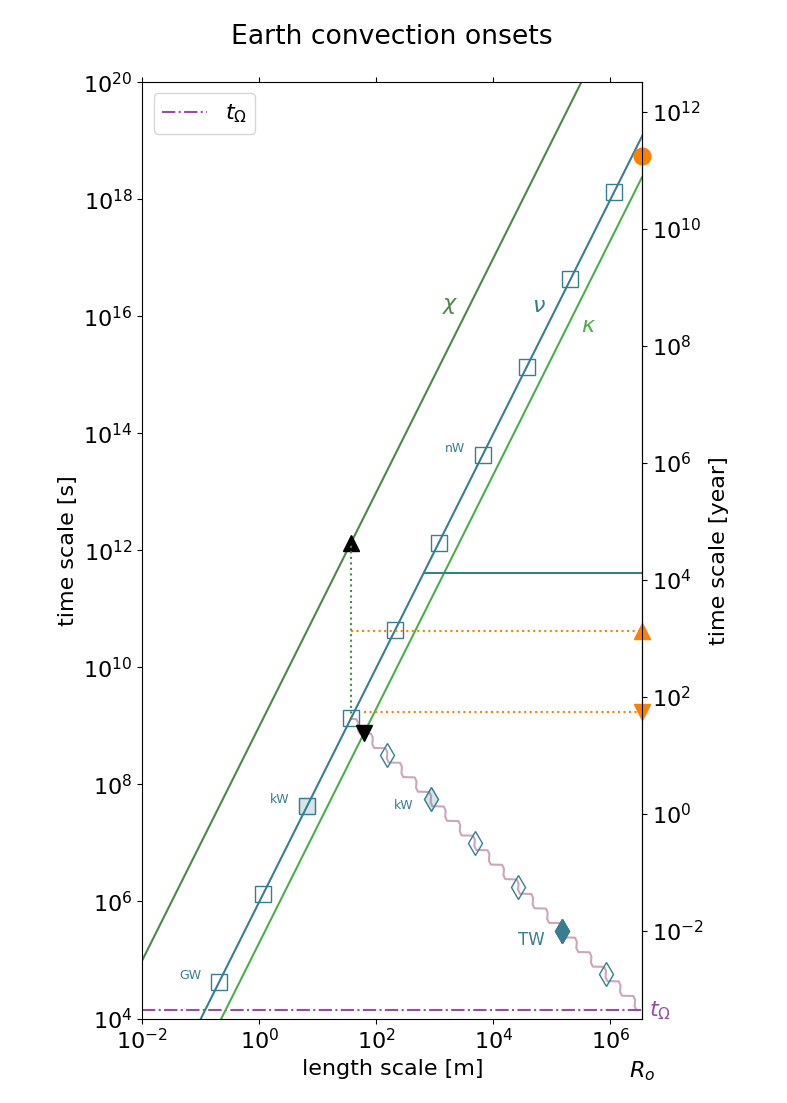}
		\caption{\tauell diagram illustrating convection onset in the Erth's core in the absence of a magnetic field.
		In the absence of rotation, convection sets in with a cell size comparable with $R_o$.
		Orange disk marks the value of $\tau_\rho$ at convection onset.
		It is half-way between $\tau_\kappa(R_o)$ and $\tau_\nu(R_o)$, where $\Ra(R_o) \sim 1$.
		Rotation controls the length-scale at which convection sets in.
		We compare a  $\Pr < 1$ case with a $\Pr > 1$ case.
		Black down-triangle marks column radius $\ell_c$ and period $\tau_c$ of the quasi-geostrophic vortices that form at convection onset when considering viscosity and thermal diffusivity ($\Pr<1$).
		Orange down-triangle at $\ell=R_o$ indicates the value of $\tau_\rho$ at onset.
		Up-triangles display the same quantities when considering viscosity and compositional diffusivity $\chi$ ($\Pr>1$). 
		Dotted lines help reading their graphical construction.
		}
		\label{fig:onset}
	\end{figure}
\end{centering}

Although \tauell regime diagrams are built to span a large range of length- and time-scales, they provide an interesting insight on what controls single scales appearing at the onset of convection.
Here we compare what would be the convection threshold in Earth's core, depending on whether it is rotating or not.
We display the results in Figure \ref{fig:onset}, using properties of the Earth's core listed in Table \ref{tab:properties}.
We extended the figure to very long $\tau$-values in order to include the intersection of the viscous line and the right $y$-axis.

In the absence of rotation, the threshold of convection is governed by a balance between buoyancy and the combined action of momentum and thermal diffusions.
It takes place at the largest length-scale $\ell = R_o$ (or at `scale height' $\mathcal{H} = \frac{C_P}{\alpha g}$ if it is smaller than $R_o$), and for $\Ra(R_o) \sim 1$ (note that the critical value $\Ra_c$ is in fact much larger than 1 due to several powers of $2\pi$, which we dropped for simplicity).
Expressing $\ell$-scale Rayleigh number as $\Ra(\ell) = \frac{\tau_\kappa(\ell) \tau_\nu(\ell)}{\tau_\rho^2(\ell)}$ as in Table \ref{tab:dimensionless}, we obtain critical $\tau_\rho$ at the convection onset: $T_\rho^c = \sqrt{\tau_\kappa(R_o) \tau_\nu(R_o)}$.
This value is plotted in Figure \ref{fig:onset} as an orange disk on the right $y$-axis, at a time half-way between $\tau_\kappa(R_o)$ and $\tau_\nu(R_o)$.

Things get very different when the system is rapidly rotating.
Proudman-Taylor constraint inhibits convective flows, and viscosity is needed to break this geostrophic constraint \citep{chandrasekhar1961}. 
Convection marginal stability in rapidly rotating spheres has a long history \citep{chandrasekhar1961,roberts1968,busse1970,jones2000,zhang2007}.
Convective structures are found to be quasi-geostrophic at onset, forming columnar vortices with a width small enough to enable viscosity in the bulk to alleviate Proudman-Taylor constraint, yielding the famous $\Ek^{1/3}$ law.
These structures can also be viewed as thermal Rossby waves.

It is interesting to examine the graphical \tauell representation of this situation.
We follow the local stability analysis of \citet{busse1970}. 
Omitting numerical prefactors, his equations (4.11) to (4.13) yield:
\begin{eqnarray}
	\ell_c & \simeq & R_o \left( \Pr^{-1} +1 \right)^{1/3} \Ek^{1/3} \\
	\tau_c & \simeq & t_\Omega \, \Pr^{1/3} (1+ \Pr)^{2/3} \Ek^{-1/3} \\
	T_\rho^c & \simeq & t_\Omega\left(  \Pr^{-1} +1\right)^{2/3} \Pr^{1/2} \Ek^{-1/3} ,
\end{eqnarray}
where $\ell_c$ is column radius, $\tau_c$ time period, $T_\rho^c$ free-fall time (for $\ell=R_o$) at onset, and $\Pr = \nu/D$ the Prandtl number, where diffusivity $D$ is either thermal diffusivity $\kappa$ or compositional diffusivity $\chi$.
Note that $\Ek$ is here the classical large-scale Ekman number $\Ek(R_o)$.

Figure \ref{fig:onset} translates these results graphically.
We compare cases with $\Pr<1$ and $\Pr>1$, by selecting diffusivity $D = \kappa$ or $D = \chi$.
At convection onset, column width is at the intersection of line $\tau_{Rossby}$ and either line $\tau_\nu$ or line $\tau_D$, depending on which it encounters first.
Thermal (or compositional) diffusivity governs period since $\tau_c \simeq \tau_D(\ell_c)$ in both cases.
We also observe that $T_\rho^c = \sqrt{\tau_\nu(\ell_c) \tau_D(\ell_c)}$.
Graphically, this places $T_\rho^c$ along the right $y$-axis ($\ell=R_o$) at a time half-way between $\tau_\nu(\ell_c)$ and $\tau_D(\ell_c)$ on our log-log plots.
Note that onset parameters from global stability analysis \citep{jones2000,dormy2004} differ substantially from those of the local stability analysis of \citet{busse1970} for $\Pr < 1$.

\bibliographystyle{crgeos}


\bibliography{../../../../biblio}

\end{document}


\maketitle

\beginsupplement
\section{\tauell Python package}
\label{sec:Python}
In the attached tau-ell\_programs.zip archive, we provide short Python scripts used to draw our article's figures.

\begin{itemize}
\item read\_parameters.py: document the parameters of the natural object, or numerical simulation, or Lab experiment, for which one wishes to draw \tauell regime diagrams (or templates). The objects can be fluid full spheres (as in the article) or fluid spherical shells.
\item plot\_object.py: plot \tauell diagram of the chosen object. You select a number of options, including the force balance you want to test, and the program calls plot\_template and plot\_scenario.
\item plot\_template.py: plot the template for the chosen object (common to natural objects, DNS and experiments).
\item plot\_scenario.py: overlay the \tauell regime diagram produced by scenarios built upon various force balances (CIA, QG-CIA, QG-VAC, MAC, QG-MAC, QG-MAC\_JA, IMAC). More scenarios can be added.
\item plot\_DNS.py: plot \tauell diagrams of numerical simulations, given their spherical harmonic degree $n$-spectra.
\item plot\_experiment.py: plot \tauell diagrams of Lab experiments, given their wavenumber $k$-spectra.
\item plot\_Kolmogorov.py: plot \tauell diagram of Kolmogorov's universal turbulence, as an example.
\item plot\_convection\_onset.py: plot dynamo onset parameters for rotating and non-rotating convection (full sphere).
\item tau\_ell\_lib.py: library gathering \tauell conversion rules from spectra, graphical functions, and other common functions.
\end{itemize}

The numerical simulation and Lab experiment data used for the examples shown in our article are available in folders:
\begin{itemize}
\item DNS\_Guervilly: $u$ and $\rho$ spectra of \citet{guervilly2019}'s 3D rotating convection simulation at $\Ek=10^{-8}$.
\item DNS\_Schaeffer: $u$, $b$, and $\rho$ average spectra of \citet{schaeffer2017}'s S2 numerical geodynamo simulation.
\item DNS\_Dormy: $u$, $b$, and $\rho$ average spectra of weak and strong dynamo numerical simulations proposed by \citet{dormy2016}.
\item experiment\_DTS: $u$ and $b$ wavenumber spectra of a composite run of the DTS liquid sodium experiment \citep[\textit{e.g.,}][]{brito2011}.
\end{itemize}

This package is also available at
\href{https://gricad-gitlab.univ-grenoble-alpes.fr/natafh/shell_tau-ell_programs}{https://gricad-gitlab.univ-grenoble-alpes.fr/natafh/shell\_tau-ell\_programs}.

\section{\tauell regime diagram of the DTS liquid sodium experiment}
\label{sec:DTS}
\begin{centering}
	\begin{figure}[h]
		\includegraphics[width=7cm]{ 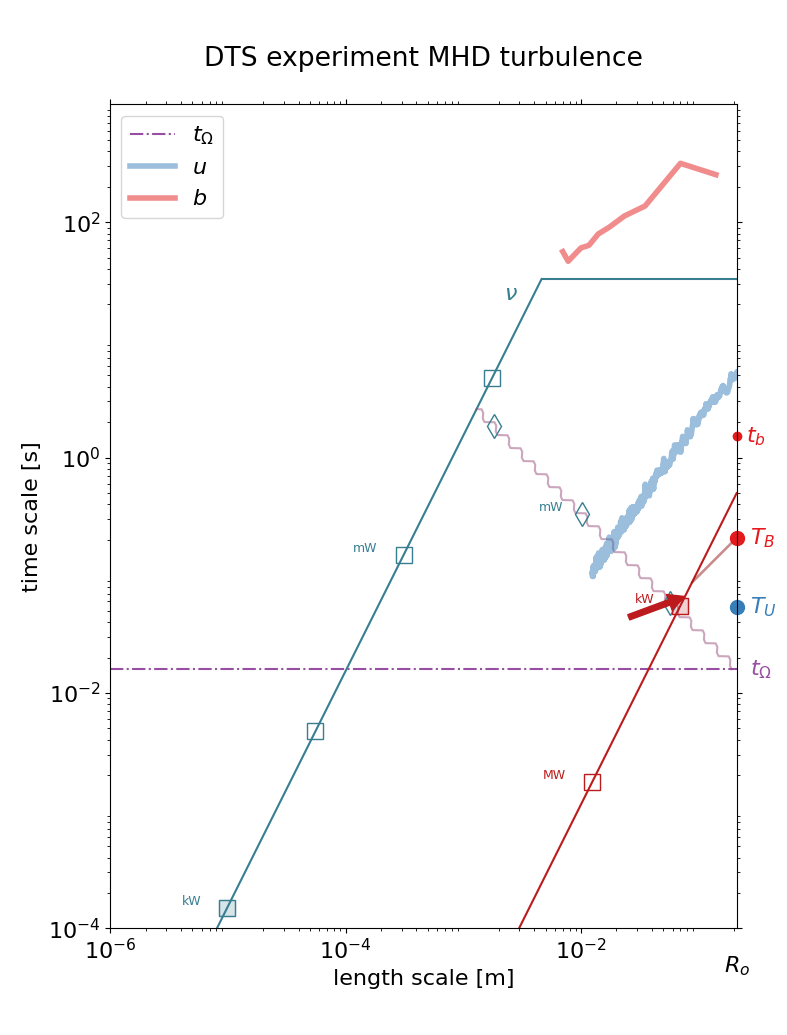}
		\caption{Composite representative \tauell diagram of DTS magnetized spherical Couette laboratory experiment.
		Outer shell rotation frequency is 10 Hz, while inner sphere spins at 20 Hz.
		Time $T_B$ (large red dot) is obtained from the energy of the dipolar magnetic field applied by the inner sphere magnet, while time $t_b$ (small red dot) represents the time-averaged induced magnetic field.
		Time $T_U$ (blue dot) is deduced from the time-averaged kinetic energy.
		Red horizontal dotted line marks dissipated power (700 W).
		Thick red line $\tau_b(\ell)$ is obtained from a $k$-spectrum of magnetic fluctuations measured at the outer shell surface.
		Thick blue line $\tau_u(\ell)$ is obtained from a frequency-spectrum of flow velocity fluctuations measured at mid-depth in the fluid using Ultrasound Doppler Velocimetry.
		}
		\label{fig:DTS}
	\end{figure}
\end{centering}

We think that laboratory experiments can also provide a better perspective when translated into \tauell regime diagrams.
As an illustration, we present in Figure \ref{fig:DTS} the composite \tauell diagram of a representative run of the DTS experiment.
The DTS experiment is a magnetized spherical Couette experiment.
Fifty liters of liquid sodium are enclosed in a spherical container ($R_o=0.21$ m) that can rotate around a vertical axis.
A central inner sphere ($R_i=0.074$ m) can rotate independently around the same axis, and hosts a strong permanent magnet producing an axial dipolar magnetic field.
See \citet{nataf2008a,brito2011} for more details.

We draw $\tau_\nu$ and $\tau_\eta$ lines from properties of liquid sodium at $130$ C ($\nu=6.5 \times 10^{-7}$ m$^2$s$^{-1}$, $\eta = 8.8 \times 10^{-2}$ m$^2$s$^{-1}$).
Power markers along these lines are deduced from liquid volume and density ($\rho = 930$ kg\,m$^{-3}$).
From the energy of applied dipolar magnetic field, we deduce time $T_B$ and draw the $\tau_{Alfven}(\ell)$ red wavy line of Alfv\'en wave propagation. 
We pick a run with outer shell rotation rate $f_o = 10$ Hz, which yields horizontal line $t_\Omega$ and wavy line $\tau_{Rossby}$.

We consider an inner sphere differential rotation rate $\Delta f = 10$ Hz, for which a power dissipation $\mathcal{P}_{diss} \simeq 700$ W is measured.
\citet{cabanes2014} reconstructed time-averaged flow and induced magnetic field by a joint inversion of a comprehensive set of flow velocity profiles and magnetic field measurements.
Their Table III provides the energies of these fields, which we convert into vortex overturn time at integral scale $T_U$, and the corresponding time for induced magnetic field $t_b$.
Note that their analysis if for $f_o = 0$, but should approximatively apply to our case.

The three components of the induced magnetic vector are measured every 6\deg along a meridian between latitudes -57\deg and +57\deg.
We thus computed a $k$-spectrum of magnetic energy density from a set of 60s-long records, which was converted into line $\tau_b(\ell)$ according to equation (42)
of the article's Appendix A.
%
No velocity measurement was available for that run, but a nice profile of angular velocity was measured using Ultrasound Doppler Velocimetry for another run with $f_o = 5$ Hz and $\Delta f = 10$ Hz.
Using equation (61) of Appendix A,
we extract the frequency power spectral density of a 40s-long record  at fluid mid-depth, and convert it to a kinetic energy density $k$-spectrum using equation (62)
with $U$ deduced from the same profile, yielding line $\tau_u(\ell)$ drawn in Figure \ref{fig:DTS}.
Note that this spectrum might be contaminated by instrumental noise.
The resulting \tauell diagram suggests that velocity fluctuations are mostly quasi-geostrophic because line $\tau_u$ plots above Rossby line.
Magnetic energy fluctuations are in a strongly dissipative region, and are almost three orders of magnitude smaller that kinetic energy fluctuations, in agreement with observations of \citet{figueroa2013}.
Under the combined influence of strong rotation and strong imposed magnetic field, energy fluctuations of both types are two to three orders of magnitude smaller than time-averaged energies (tagged by $T_U$ and $t_b$ in Figure \ref{fig:DTS}), as noted by \citet{nataf2008}, and \citet{kaplan2018}.

The short red wavy line $\tau_{Alfven}(\ell)$ indicates that geostrophic Alfv\'en waves can propagate but are severely damped, as analyzed by \citet{tigrine2019}.
Dissipation is dominated by Ohmic dissipation of the time-averaged flow.

\bibliographystyle{crgeos}


\bibliography{../../../../biblio}